\documentclass[sigconf]{acmart}

\usepackage{makecell}
\usepackage{multirow}
\usepackage{longtable}
\usepackage{graphicx}
\usepackage{caption}
\usepackage{subcaption}
\usepackage{tabularx}
\usepackage{soul}
\usepackage{float}
\usepackage{url}
\usepackage{booktabs}
\usepackage{colortbl}

\usepackage{enumitem}

\usepackage[most]{tcolorbox}
\usepackage{xcolor}

\definecolor{sketchspeech}{HTML}{EE8A63}
\definecolor{aiinterpret}{HTML}{5B8FD9}
\definecolor{humanfollowup}{HTML}{D95F86}
\definecolor{laterinteraction}{HTML}{D6A62E}

\definecolor{c1color}{HTML}{EE8A63}
\definecolor{c2color}{HTML}{D95F86}
\definecolor{c3color}{HTML}{6B7280}

\newcommand{\figblock}[2]{%
  \tcbox[on line,colback=#1!6,colframe=#1!85!black,
    coltext=black,boxrule=.6pt,arc=2pt,boxsep=0pt,
    left=4pt,right=4pt,top=2pt,bottom=2pt,
    fontupper=\normalfont\fontsize{8}{9.5}\selectfont]{#2}%
}
\DeclareRobustCommand{\designintentcode}{\figblock{c1color}{Design Intent}}
\DeclareRobustCommand{\followupcode}{\figblock{c2color}{Human Follow-up}}
\DeclareRobustCommand{\agencycode}{\figblock{c3color}{Creative Agency}}

\newif\ifredact
\redactfalse  

\newcommand{\ts}{\textit{CommSketch}}
\newcommand{\bl}{\textit{Sketch-Only}}
\newcommand{\dquote}[1]{\textit{``#1''}}

\newif\ifcomment
\commenttrue

\ifcomment
  \newcommand{\missing}[1]{{\color{red}#1}}
  \newcommand{\wei}[1]{{\sethlcolor{cyan!40}\hl{[Weiyan: #1]}}}
  \newcommand{\ken}[1]{{\sethlcolor{yellow!40}\hl{[Kenny: #1]}}}
  \newcommand{\ger}[1]{{\sethlcolor{green!40}\hl{[Geraldine: #1]}}}
  \newcommand{\rrev}[3]{{\color{blue}[RevID #1] #2: #3}}
\else
  \newcommand{\missing}[1]{}
  \newcommand{\wei}[1]{}
  \newcommand{\ken}[1]{}
  \newcommand{\ger}[1]{}
  \newcommand{\rrev}[3]{}
\fi

\AtBeginDocument{%
  }

\setcopyright{acmlicensed}
\copyrightyear{2018}
\acmYear{2018}
\acmDOI{XXXXXXX.XXXXXXX}
\acmConference[Conference acronym 'XX]{Make sure to enter the correct
  conference title from your rights confirmation email}{June 03--05,
  2018}{Woodstock, NY}
\acmISBN{978-1-4503-XXXX-X/2018/06}

\begin{document}

\title{CommSketch: How Speaking while Sketching Steers Human--AI Design Ideation}
\author{Weiyan Shi}
\email{weiyanshi@gmail.com}
\orcid{0009-0001-6035-9678}
\affiliation{
  \institution{Singapore University of Technology and Design}
  \city{Singapore}
  \country{Singapore}
}

\author{Darryl Lim}
\email{owlycodes@gmail.com}
\orcid{0009-0007-8173-2199}
\affiliation{
  \institution{Singapore University of Technology and Design}
  \city{Singapore}
  \country{Singapore}
}

\author{Geraldine Quek}
\email{geraldine_quek@sutd.edu.sg}
\orcid{0000-0003-2864-3860}
\affiliation{
  \institution{Singapore University of Technology and Design}
  \city{Singapore}
  \country{Singapore}
}

\author{Kenny Tsu Wei Choo}
\email{kennytwchoo@gmail.com}
\orcid{0000-0003-3845-9143}
\affiliation{
  \institution{Singapore University of Technology and Design}
  \city{Singapore}
  \country{Singapore}
}

\ccsdesc[500]{Human-centered computing~Interactive systems and tools}
\ccsdesc[500]{Human-centered computing~Human computer interaction (HCI)}

\keywords{generative AI, human--AI co-creativity, design ideation, sketching, speech interaction, multimodal interaction, design intent, creative agency}

\begin{teaserfigure}
  \centering
  \includegraphics[width=0.7\textwidth,height=0.22\textheight,keepaspectratio]{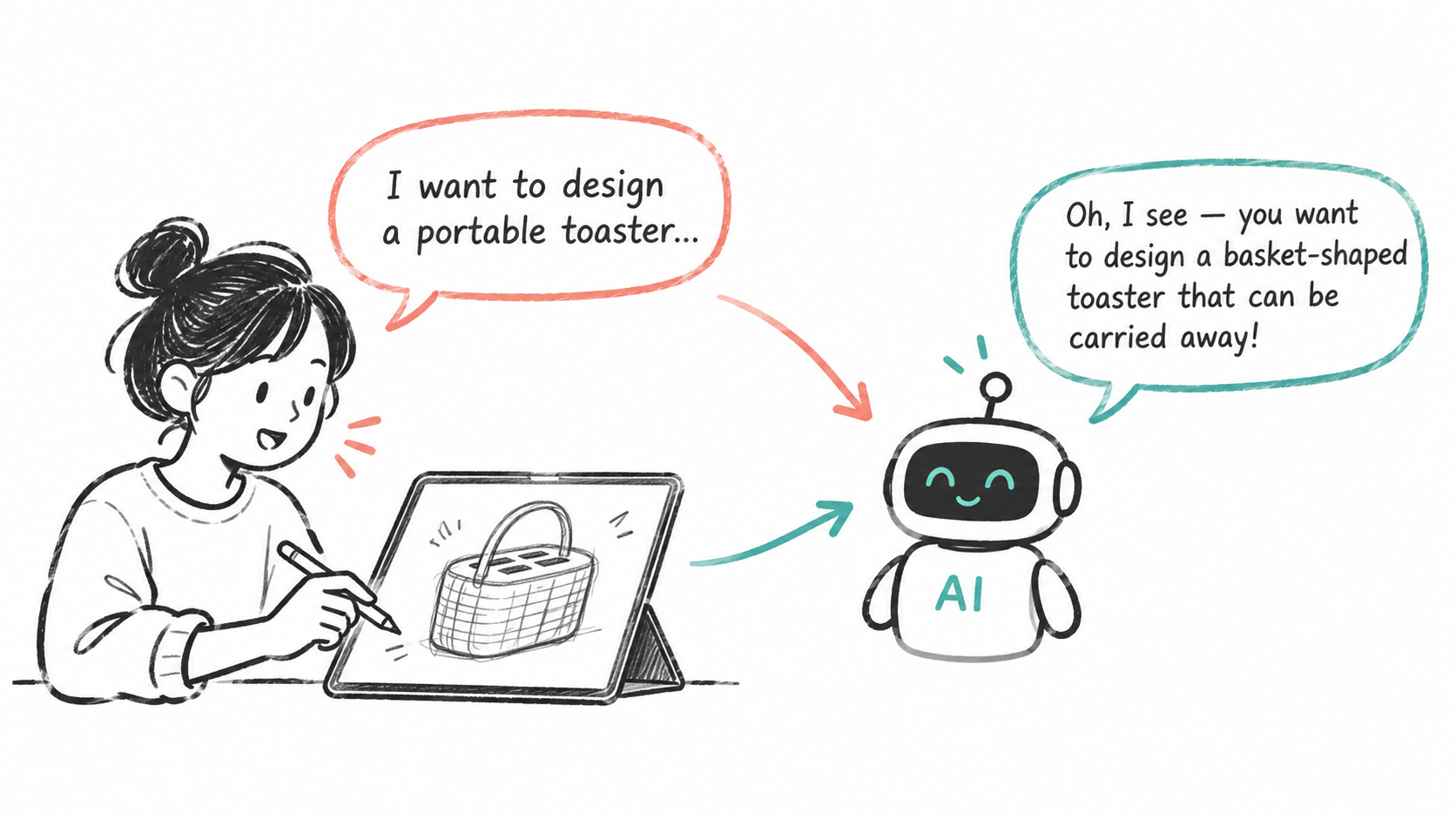}
  \caption{\ts{}: Communicating design intent through sketch and speech.
  A designer sketches a basket-shaped toaster while describing its intended
  portability; the AI combines the visual form and spoken intention to
  interpret the emerging design.}
  \Description{Conceptual illustration of a designer drawing a basket-shaped
  toaster on a tablet while saying they want a portable toaster. Arrows from
  the sketch and speech lead to an AI character, whose response connects the
  basket-shaped form with the intention that the toaster can be carried.}
  \label{fig:banner}
\end{teaserfigure}

\begin{abstract}
Designers often speak while sketching when explaining ideas, yet AI design tools often rely on sketches or prompts, overlooking context expressed as ideas develop. We developed a sketch-based AI design interface that jointly interprets sketches and concurrent speech. Through a between-subjects study ($N=24$), we examined how speaking while sketching steers human--AI design ideation compared with sketches alone. For creativity support, concurrent speech supported natural expression of design intent and efficient visualisation. For human--AI collaboration, speech helped establish a shared understanding of design intent, supported significantly higher perceived alignment ($p<.05$), and enabled participants to guide AI contributions as ideas co-evolved. We discuss how future human--AI design tools could support dynamic alignment, broader multimodal expression, and human--AI co-creativity.
\end{abstract}

\maketitle

\section{Introduction}

Recent advances in generative AI have created new opportunities to support
design ideation, helping designers explore, elaborate, and visualise ideas
as they emerge. Sketching remains central to this process: sketches are
quick to produce, easy to change, and expressive enough to suggest form,
layout, and spatial relationships before a design is fully
specified~\cite{buxton2010sketching,goldschmidt1991dialectics}.
This openness makes sketching especially valuable during ideation, when
designers may still be discovering not only how an idea should look, but
what they want to make~\cite{suwa2022roles}.

However, the same openness that makes sketches useful for ideation also
makes them difficult for AI to interpret. Early sketches are often rough,
incomplete, and ambiguous~\cite{purcell1998drawings}, and their marks capture
only part of the designer's evolving intent. A few lines may suggest an
object's form without conveying what it represents, how it should work,
or why particular features matter for intended experiences such as
portability or ease of use. When AI design tools rely on sketches alone,
they must infer this context, potentially producing interpretations that
diverge from the designer's intent and requiring designers to explain,
correct, or reframe their ideas to restore alignment.

Current AI-supported design tools often address this problem by asking
designers to provide clearer input. Designers may add a prompt, upload a
reference image, or describe what they want before requesting a
result~\cite{peng2024designprompt,tao2025designweaver,lin2025inkspire,davis2025sketchai}.
These approaches can help AI tools better understand design intent, but
often frame communication with AI as a separate step from sketching.
The designer first draws, then translates the idea into a more explicit
request, or later corrects the AI after it has already misunderstood the
sketch~\cite{shi2026talksketch,cheng2025aiawareness}.

This separation overlooks how designers often communicate ideas in human
design practice. When explaining a sketch to another person, designers
commonly speak as they draw~\cite{kim2009study,purcell1998drawings}.
They may name an emerging object, describe how it should behave, clarify
an ambiguous mark, or explain why they are changing part of the design.
In these moments, speech is not merely a caption added to a finished
drawing. It is part of how the sketch becomes meaningful while the idea
is still taking shape.

Speaking while sketching may therefore offer a different way to support
human--AI design ideation. Rather than requiring designers to package
intent into a prompt before or after sketching, concurrent speech can
make intent available as the sketch develops. This could help AI
understand not only the visible marks, but also the designer's evolving
understanding of what those marks mean. It may also keep the interaction
closer to the natural flow of ideation, allowing designers to express
intent for AI understanding while continuing to sketch.

However, concurrent speech may contain tentative possibilities,
self-reflections, or ideas that are later abandoned, rather than fixed
instructions. Using such speech as design context raises questions about
how AI interpretations influence what designers subsequently clarify,
revise, or develop. Understanding its role therefore requires examining
the interaction process through which human and AI contributions unfold.

Prior work has explored ways to combine sketch, speech, and prompts for
AI interaction~\cite{rosenberg2024drawtalking,huang2025sketchgpt,shi2026drawing}.
However, less is known about how concurrent speech shapes the unfolding
process of human--AI design ideation and designers' perceptions of
creativity support and collaboration.

Compared with grounding AI in sketches alone, we ask:

\textbf{RQ1. How does speaking while sketching steer human--AI
design ideation?}

\textbf{RQ2. How does this form of human--AI interaction shape designers'
perceptions of creativity support and human--AI collaboration?}

We investigated these questions through a sketch-based AI interface
that jointly interprets sketches and concurrent speech.
Our contributions are threefold:

\begin{itemize}
\item \textbf{A comparative study of speaking while sketching:}
we conducted a between-subjects study with 24 participants to investigate
how grounding AI in sketches and concurrent speech, compared with sketches
alone, shapes design ideation and designers' perceptions of creativity
support and human--AI collaboration.

\item \textbf{An empirical account of how speaking while sketching steers
human--AI ideation:}
we examine how speech supports natural expression of design intent and
efficient visualisation, contributes to perceived alignment, and shapes
how participants guide and develop AI contributions.

\item \textbf{Design implications for future human--AI design tools:}
we identify opportunities to support dynamic alignment with evolving
intent, broaden multimodal expression, and support human--AI co-creativity
through speech.
\end{itemize}
\section{Related Work}

We review three strands of work that situate speaking while sketching within human--AI design ideation: multimodal approaches to communicating user intent, creativity tools that incorporate sketches into generative workflows, and research combining speech and sketching. Across these strands, we consider how emerging intentions become available to AI and how this informs the study of subsequent interaction.

\subsection{Multimodal Intent Communication in Human--AI Interaction}

Conversational User Interfaces (CUIs) enable dialogue-based interaction that resembles human conversation~\cite{mctear2002spoken} and are widely deployed through chatbots~\cite{folstad2017chatbots} and voice-based assistants~\cite{de2020intelligent}. Recent advances in generative AI (GenAI), including large language models (LLMs)~\cite{achiam2023gpt}, multimodal assistants~\cite{team2024gemini}, and image-generation systems~\cite{rombach2022high}, have expanded the range of tasks that can be performed through conversational interaction. These systems have been applied across domains including education, healthcare, creative coding, and collaborative ideation~\cite{shaer2024ai,angert2023spellburst,ramjee2025ashabot}.

Despite this flexibility, communicating intent to generative AI remains difficult. Users often need to translate incomplete or evolving goals into explicit prompts before a system can act. Prior work has therefore explored multimodal and interface-based approaches that make additional aspects of user activity available to AI. For example, GesPrompt combines speech with co-speech gestures to support intent communication in extended-reality environments~\cite{hu2025gesprompt}, while Persistent Assistant incorporates embodied input and multimodal feedback into everyday AI interaction~\cite{cho2025persistent}. Other systems embed LLMs within direct-manipulation interfaces. DirectGPT~\cite{masson2024directgpt} and Spellburst~\cite{angert2023spellburst}, for instance, translate interface actions and graphical structures into information that can guide model responses. Related studies have examined how users iteratively express and refine intent in creative coding, data analysis, and immersive authoring~\cite{subramonyam2024bridging,weng2025insightlens,zhang2024vrcopilot}.

Together, these approaches show how AI can draw on speech, gesture, interface actions, and visible artefacts to interpret user intent. Sketch-based ideation offers a setting in which to examine these resources while both the visual representation and the designer's intention are still developing. A rough mark may suggest a form without specifying its identity, intended use, or relation to other features. The challenge is to make this emerging meaning available for interaction while preserving the openness that makes sketching useful.

\subsection{AI-Supported Early-Stage Design Ideation}

Sketching is a foundational practice in early-stage design, supporting rapid externalisation, exploration, reflection, and communication of ideas~\cite{buxton2010sketching,goldschmidt1991dialectics}. Sketches can communicate visual form and spatial relationships while also functioning as cognitive artefacts through which designers develop and reconsider emerging concepts~\cite{tversky2013visualizing,fan2023drawing}. Their value lies partly in their incompleteness: rough representations allow ideas to remain open to reinterpretation and change.

Generative AI has increasingly been incorporated into design tools to support visual exploration, layout generation, and early prototyping. Commercial platforms such as Adobe Firefly, Canva, Midjourney, Uizard, and Figma provide prompt-based mechanisms for generating and refining visual artefacts~\cite{adobefirefly2025tools,canva2025overview,midjourney2025design,uizard2025ai,figmaai2025update}. Research prototypes have explored richer forms of designer-authored input. DesignPrompt enables designers to compose text, images, and colours to communicate intent~\cite{peng2024designprompt}, while DesignWeaver exposes latent design dimensions from generated images to support structured refinement~\cite{tao2025designweaver}.

Other systems place sketching more directly within the generative workflow. Inkspire combines analogical sketching with generative feedback to support design exploration~\cite{lin2025inkspire}, while SketchAI adopts a sketch-first approach to AI-supported fashion design~\cite{davis2025sketchai}. These systems demonstrate the value of preserving sketching as part of interaction with generative AI rather than replacing it with text prompting alone.

These workflows make designer-authored inputs available for generation and refinement. During early ideation, however, the meaning of a sketch may also emerge through what designers say while producing it. Designers may name a feature, consider alternatives, or explain a change while still deciding what to draw. Such expression can provide context for a submitted sketch even when it has not been formulated as a request. This raises the question of how concurrent expression informs AI interpretation and the subsequent development of an idea.

\subsection{Speaking While Sketching in Human--AI Ideation}
\label{sec:rw-speech}

Sketching and speech provide complementary resources for expressing and developing design intent. Sketches are particularly effective for representing visual form, proportion, and spatial organisation, whereas speech can communicate object identities, functions, rationales, intended experiences, and contextual requirements that may be difficult to depict visually. Speech has long been recognised as a resource for reasoning, problem solving, and interaction with graphical systems~\cite{winograd1972understanding,bolt1980put,fry1979physics}. In design activity, verbalisation may also accompany the development of an idea, allowing designers to name emerging elements, explain decisions, consider alternatives, and evaluate what they have drawn.

Existing interactive systems have often treated speech as an explicit control channel. Spoken language may be used to issue commands, dictate prompts, label objects, or specify properties after or alongside a visual action. Early sketch-based AI systems similarly tended to frame speech as supplementary input supplied intentionally to help the system interpret a sketch, rather than as verbalisation produced through the ongoing process of sketching~\cite{duan2026sketchconcept}. This distinction matters because speech produced while sketching may not be fully planned or directed towards the AI. It can be tentative, fragmented, self-directed, exploratory, or revised as the design develops.

Recent systems have begun to support sketching and speaking concurrently to capture richer user intent. For instance, DrawTalking allows users to verbalise object identities, properties, and behaviours while constructing interactive sketches, and explores creative uses through an open-ended study~\cite{rosenberg2024drawtalking}, while StoryDrawer transforms children's real-time spoken narratives into complementary sketches~\cite{zhang2022storydrawer}. Similarly, SketchGPT unifies concurrent sketch and speech to communicate intent to large language models, comparing multimodal interaction with speech-only and sketch-only alternatives~\cite{huang2025sketchgpt}. Focusing on sketch-to-image generation, Shi et al. empirically showed that incorporating concurrent speech significantly improves the alignment between generated design images and designers' stated intentions~\cite{shi2026drawing}. Together, these works demonstrate that concurrent speech provides critical semantic context beyond what is visually represented, thereby enhancing intent interpretation and generation fidelity.

These studies address complementary questions about interactive authoring, multimodal intent expression, and the alignment of generated outputs. They also offer evidence about users' experiences with speech and sketch interfaces. Building on these findings, our focus is how making concurrent speech available to AI relates to participants' subsequent responses and the development of design ideas.

Exploratory qualitative research provides initial evidence for this broader perspective. In a formative study ($N=6$), Shi et al. found that designers developed ideas through ongoing sketching but experienced friction when required to interrupt this activity and formulate lengthy textual prompts~\cite{shi2026talksketch}. Participants expected AI to remain aware of their developing sketches and verbalised thinking, rather than requiring each interaction to begin with a self-contained request. Similarly, Cheng et al.'s Wizard-of-Oz study ($N=20$) found that limited AI awareness of designers' speech, gaze, pen strokes, and evolving sketch state increased interaction friction and reduced communication efficiency~\cite{cheng2025aiawareness}. These studies suggest that speech may form part of the continuous activity through which design intent is externalised, rather than simply functioning as a separate prompting modality.

Building on this work, we examine how speaking while sketching steers the interaction process and shapes perceived creativity support and human--AI collaboration. Using a functioning AI system, we examine how designers communicate emerging intent and respond to and develop AI contributions during ideation.

\section{Method}

\subsection{Study Design}

We conducted a between-subjects study to investigate the role of speaking while
sketching in canvas-based human--AI design ideation. The study compared two
conditions: (1) \bl{}, in which participants sketched their ideas and the AI
interpreted the current design from the sketch alone; and (2) \ts{}, in which
participants could speak while sketching and the AI interpreted the current
design from the sketch and any concurrent speech. Participants chose how to
use the available functions; no think-aloud protocol was imposed.

Both conditions used the same prototype, design task, interaction workflow,
chatbot functions, AI model, prompting strategy, and generation pipeline. Thus,
the manipulation focused on whether verbalisation produced during sketching was
captured as part of the ongoing design activity and used as additional
grounding for the AI's interpretation. This allowed us to examine how speaking
while sketching influenced participants' experience of creativity support and
human--AI collaboration, their uptake of AI interpretations, and the resulting
patterns of creative agency during ideation.

\subsection{Design Rationales}

The prototype was guided by four design rationales, drawing on prior work
on sketch-based human--AI ideation and multimodal interaction.

\subsubsection{Low-Friction Human--AI Ideation}
We embedded AI support directly within the sketching environment so that
designers could move between developing ideas and engaging with AI without
switching tools or repeatedly reformulating evolving ideas as standalone
prompts. This design aimed to keep AI support available throughout ideation,
with designers able to return to drawing after engaging with the assistant
and carry developing ideas across these activities
\cite{shi2026talksketch,shi2026taxonomy,lin2025inkspire,lawton2023drawing}.

\subsubsection{Seamless Capture of Speaking While Sketching}
In \ts{}, participants' concurrent verbalisation was recorded and
transcribed in the background while they sketched. They could express
developing ideas without activating a voice-command mode or formulating
complete instructions to AI. A peripheral recording indicator made the
recording state visible while keeping attention on the canvas.
This design treated speech as an additional source of context for the
evolving sketch~\cite{oviatt1999ten,turk2014multimodal,shi2026talksketch}.
Recording paused during chatbot interaction and resumed on returning to
the canvas, keeping the captured verbalisation tied to the sketching activity.

\subsubsection{Inspectable AI Understanding of Design}
We used checkpoint-based interpretation: opening the chatbot triggered a
brief AI interpretation of the current design. The system did not infer or
update design interpretations while participants were sketching. In the \bl{} condition, this
interpretation was based on the sketch alone. In the \ts{}
condition, it was based on both the sketch and the transcript of speech
produced while sketching. This made the AI's understanding visible before
participants submitted an explicit prompt. Participants could confirm,
correct, question, elaborate on, or build upon that interpretation.
Making this understanding inspectable provided a basis for participants
to identify misunderstandings and communicate further intentions
\cite{cho2025persistent,chen2025need,shi2026talksketch}.

\subsubsection{Generative Support for Idea Development}
The chatbot supported text and image generation so that participants could
explore, extend, and visualise their developing design ideas. Generated
outcomes could be discussed through the chatbot or imported onto the canvas
for further development. These functions connected conversational exploration
with visual development on the canvas, allowing participants to incorporate
AI-generated content into subsequent sketching
\cite{lin2025inkspire,lawton2023drawing}.

\subsection{Prototype}

We developed a sketch-based prototype comprising a digital canvas and an AI
chatbot that interpreted the current design and supported text and image
generation. Participants could move between these modules throughout idea
development. Both conditions used the same interface and generative
functions; \ts{} additionally captured speech during canvas activity and
included its transcript in the AI's interpretation of the sketch.
Figure~\ref{fig:prototype} shows the interface and shared interaction workflow.
\newsavebox{\prototypeUIbox}
\newlength{\prototypePanelHeight}
\begin{figure*}[t]
\centering
\sbox{\prototypeUIbox}{\includegraphics[width=.56\textwidth]{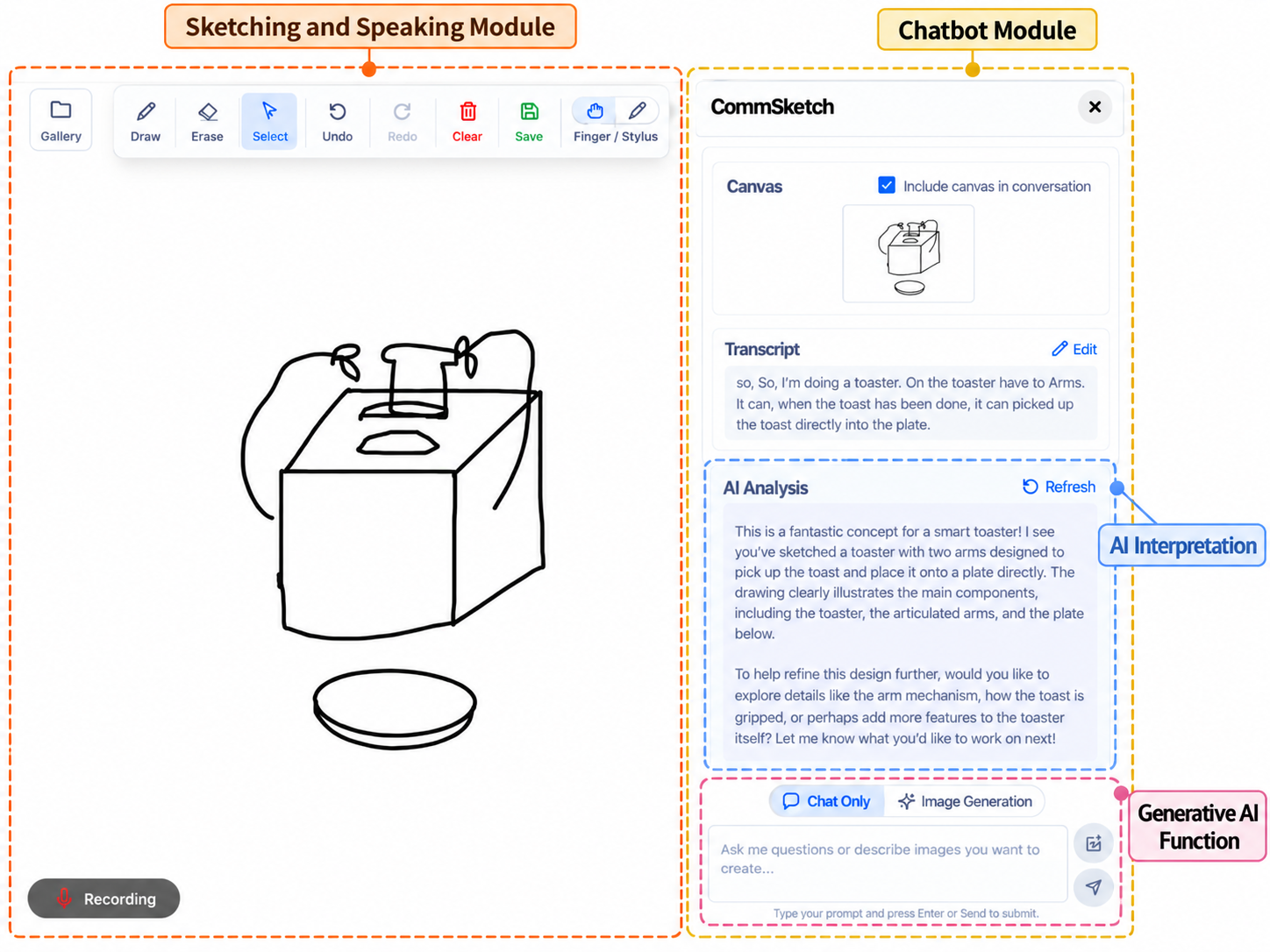}}
\setlength{\prototypePanelHeight}{\dimexpr\ht\prototypeUIbox+\dp\prototypeUIbox\relax}
\begin{subfigure}[t]{.56\textwidth}
  \vspace{0pt}
  \centering
  \usebox{\prototypeUIbox}
  \caption{Prototype interface.}
  \label{fig:prototype-ui}
\end{subfigure}\hfill
\begin{subfigure}[t]{.41\textwidth}
  \vspace{0pt}
  \centering
  \includegraphics[height=\prototypePanelHeight]{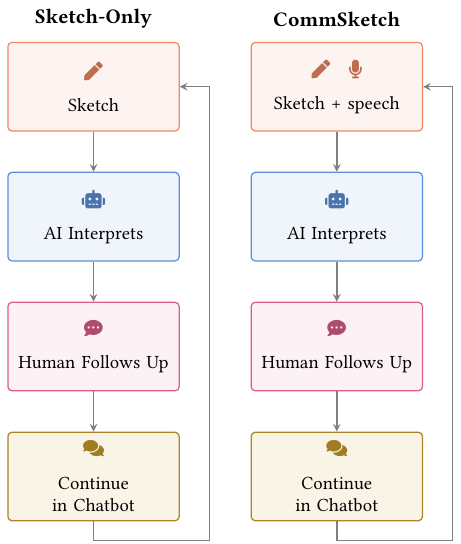}
  \caption{Condition-specific workflows.}
  \label{fig:prototype-workflow}
\end{subfigure}
\caption{Prototype interface and interaction workflow.
(a) The interface combines a sketching canvas with a chatbot for inspecting
AI interpretations and requesting text or image generation. The screenshot
shows \ts{}, which additionally captures concurrent speech.
(b) Opening the chatbot pauses canvas activity and presents an AI
interpretation grounded in the sketch alone in \bl{}, or the sketch and
speech transcript in \ts{}. Participants can respond, continue the chatbot
exchange, or return to the canvas to develop the design further, optionally
importing generated images. In \ts{}, returning to the canvas resumes
speech capture.}
\label{fig:prototype}
\Description{Left: the complete annotated CommSketch interface, showing the
sketch canvas, speech transcript, AI interpretation, and generative controls.
Right: two separate workflows for Sketch-Only and CommSketch each show
canvas activity, AI Interprets, Human Follows Up, and Continue in Chatbot.
Each workflow proceeds down four stages and has a single return loop from
Continue in Chatbot to canvas activity. There is no return arrow from Continue
in Chatbot to Human Follows Up. All text is black; canvas icons are orange,
AI icons blue, and follow-up icons pink.}
\end{figure*}

\subsubsection{Sketching and Speaking Module}
Participants interacted with the system using an Apple iPad Pro 13" and Apple Pencil. The canvas, built with Fabric.js, supported freehand drawing, erasing, selection, undo/redo, and canvas reset using stylus or touch input. Participants could select a region of the canvas and send it to the chatbot as visual context for AI interaction.

In \ts{}, speech was automatically recorded through the iPad's built-in microphone while participants worked on the canvas and the chatbot was closed, with a peripheral indicator showing the recording state. Speech was transcribed using the Google Cloud Speech-to-Text v1 streaming API (\texttt{en-US}, automatic punctuation). Opening the chatbot stopped the recording and made the captured transcript available as context for AI interpretation. Speech capture itself did not continuously trigger AI interpretation while participants were sketching.

\subsubsection{Chatbot Module}
The chatbot maintained a shared conversation context across AI interpretation, text generation, and image generation, such that each interaction could build on the preceding chatbot history.

\paragraph{AI Interpretation}
Opening the chatbot triggered an interpretation request to \texttt{gemini-2.5-flash}; the response was displayed before participants submitted a prompt. The request contained the current sketch in \bl{}, and the current sketch together with the concurrent speech transcript in \ts{}. The interpretation prompt was deliberately simple: \dquote{Based on the user's sketch and [concurrent speech transcript], provide a brief interpretation to support further interaction.} We kept this prompt minimal to reduce the AI's influence on participants' subsequent design decisions. The sketch and transcript were supplied as aggregate context, without stroke-level timestamp alignment between speech segments and drawing actions before submission. Interpretation occurred at these user-triggered checkpoints, not continuously during canvas activity.

\paragraph{Text and Image Generation}
The chatbot accepted natural-language prompts and supported both text and image generation using Gemini 2.5 Flash and Gemini 2.5 Flash Image, respectively. Generated images could be imported directly onto the canvas.

\subsubsection{Shared Interaction Workflow}
Participants could switch between the canvas and chatbot. Opening the
chatbot disabled sketching and paused speech recording in \ts{}; returning
to the canvas restored sketching and, in \ts{}, speech capture.

\subsection{Participants}
We recruited 24 participants with prior experience in design ideation through
university coursework or professional practice (Table~\ref{tab:participants}).
Participants were randomly assigned to \bl{} ($n=12$) or \ts{} ($n=12$).
Participants received approximately USD~7.8. The study was approved by the
Institutional Review Board of [REDACTED] (Approval No. [REDACTED]).

\begin{table*}[t]
\caption{Participant demographics and self-reported frequency of AI use for design. Undergraduate (Design Coursework) denotes students outside a design programme or without a declared major who had completed design-related coursework. Frequencies are reported as times per week; C9 reported that AI use for design varied with the week's work.}
\label{tab:participants}
\centering
\begingroup
\footnotesize
\setlength{\tabcolsep}{6pt}
\renewcommand{\arraystretch}{1.12}
\begin{tabular}{@{}llllr@{}}
\toprule
ID & Gender & Age & Background & \makecell[r]{AI use for design\\(times/week)} \\
\midrule
\rowcolor[HTML]{E7EFF7}
\multicolumn{5}{@{}l}{\bl{} ($n=12$)} \\
S1 & Male & 25--30 & Architecture Master's Student & $>10$ \\
S2 & Male & 21--24 & Architecture Undergraduate & $>10$ \\
S3 & Male & 25--30 & Automotive Designer & $>10$ \\
S4 & Female & 18--20 & Undergraduate (Design Coursework) & $<5$ \\
S5 & Male & 21--24 & Engineering Design Undergraduate & 5--10 \\
S6 & Male & 21--24 & Architecture Undergraduate & 5--10 \\
S7 & Female & 18--20 & Undergraduate (Design Coursework) & $<5$ \\
S8 & Female & 18--20 & Engineering Design Undergraduate & $<5$ \\
S9 & Male & 21--24 & Undergraduate (Design Coursework) & $<5$ \\
S10 & Male & 18--20 & Undergraduate (Design Coursework) & 5--10 \\
S11 & Male & 25--30 & Human-Centred Design Master's Student & 5--10 \\
S12 & Female & 35--44 & HCI PhD Student (Former UI Lecturer) & $<5$ \\
\addlinespace[4pt]
\rowcolor[HTML]{F8EBDD}
\multicolumn{5}{@{}l}{\ts{} ($n=12$)} \\
C1 & Male & 18--20 & Engineering Design Undergraduate & 5--10 \\
C2 & Male & 21--24 & Architecture Undergraduate & 5--10 \\
C3 & Male & 21--24 & Undergraduate (Design Coursework) & 5--10 \\
C4 & Male & 21--24 & Undergraduate (Design Coursework) & 5--10 \\
C5 & Male & 18--20 & Undergraduate (Design Coursework) & 5--10 \\
C6 & Male & 18--20 & Architecture Undergraduate & $<5$ \\
C7 & Male & 18--20 & Engineering Design Undergraduate & $>10$ \\
C8 & Female & 18--20 & Architecture Undergraduate & $<5$ \\
C9 & Female & 25--30 & Architecture Master's Student (Former Architect) & $>10$ \\
C10 & Male & 21--24 & Undergraduate (Design Coursework) & $<5$ \\
C11 & Female & 25--30 & UI Designer & 5--10 \\
C12 & Female & 21--24 & Architecture Undergraduate & $<5$ \\
\bottomrule
\end{tabular}
\endgroup
\end{table*}

\subsection{User Study}
Each session lasted approximately one hour and followed four stages
(Table~\ref{tab:user-study}): a 10-minute introduction and tutorial,
a 30-minute toaster design task, 5 minutes for post-task questionnaires,
and a 15-minute retrospective interview.

\begin{table}[t]
\caption{User study stages, activities, and collected data. Participants chose how to use the available functions during ideation, including when and what to say in \ts{}.}
\label{tab:user-study}
\centering
\begingroup
\footnotesize
\setlength{\tabcolsep}{6pt}
\renewcommand{\arraystretch}{1.2}
\begin{tabularx}{\linewidth}{@{}p{0.22\linewidth}p{0.07\linewidth}XX@{}}
\toprule
Stage & Time & Activities & Collected data \\
\midrule
\cellcolor[HTML]{E6EFEB}Introduction and tutorial & 10 min & Informed consent; demographic information; introduction to functions; prototype practice. & Participant demographics and frequency of AI use for design. \\
\addlinespace[4pt]
\cellcolor[HTML]{E7EFF7}Design ideation & 30 min & Freely ideate toaster designs using \bl{} or \ts{}. & Logged human--AI interaction sequences. \\
\addlinespace[4pt]
\cellcolor[HTML]{F6EDDC}Post-task questionnaires & 5 min & Rate perceived creativity support~\cite{cherry2014quantifying,lin2025inkspire,lin2026visuallyrics} and human--AI collaboration~\cite{lawton2023drawing,lin2025inkspire}. & Seven-point item ratings (Section~\ref{sec:study-questionnaires}). \\
\addlinespace[4pt]
\cellcolor[HTML]{F1E7EB}Retrospective interview & 15 min & Review interaction history; confirm idea boundaries; reflect on experience. & Interview transcripts and confirmed idea boundaries. \\
\bottomrule
\end{tabularx}
\endgroup
\end{table}

First, participants were introduced to the study, provided informed consent,
and reported their demographic information and frequency of AI use for design. They
received a tutorial on the prototype's available functions: sketching,
opening the chatbot, reviewing AI Interprets, requesting generated content,
and importing generated images onto the canvas.

Second, all participants were asked to \dquote{freely ideate toaster designs
within 30 minutes}. They chose how to use the available functions while
developing their ideas. In \bl{}, AI Interprets used the sketch alone.
In \ts{}, speaking while sketching was available as an additional input
channel, and any concurrent speech was included alongside the sketch.
Participants chose when and what to say while using the speech functionality.
The prototype automatically logged interaction events throughout the task.

Third, participants completed adapted questionnaires on perceived creativity
support~\cite{cherry2014quantifying,lin2025inkspire,lin2026visuallyrics} and
human--AI collaboration~\cite{lawton2023drawing,lin2025inkspire}.
The measures and analysis are detailed in
Section~\ref{sec:data-collection-analysis}.

Finally, participants reviewed their interaction history with the researcher
in a semi-structured retrospective interview. They identified and confirmed
the boundaries between successive design ideas and reflected on how they
communicated ideas, responded to AI Interprets, and used generated outputs.
The interview also explored useful and limiting aspects of the interaction
and, in \ts{}, participants' experience of the speech functionality.

\subsection{Data Collection and Analysis}
\label{sec:data-collection-analysis}
We collected post-task questionnaire responses, logged human--AI interaction
sequences, and semi-structured interviews. We analysed these complementary
sources through participant-level questionnaire comparisons, interaction
sequence coding, and directed qualitative content analysis~\cite{hsieh2005three}
(Table~\ref{tab:analysis-overview}).

\begin{table*}[t]
\caption{Inputs, methods, and outputs of the three analyses. Section references point to the detailed procedures. For Design Intent, \bl{} coding used sketches and annotations; \ts{} coding additionally included concurrent speech.}
\label{tab:analysis-overview}
\centering
\begingroup
\footnotesize
\setlength{\tabcolsep}{5pt}
\renewcommand{\arraystretch}{1.2}
\begin{tabularx}{\linewidth}{@{}>{\raggedright\arraybackslash}p{0.16\linewidth}>{\raggedright\arraybackslash}p{0.18\linewidth}>{\raggedright\arraybackslash}X>{\raggedright\arraybackslash}X@{}}
\toprule
Analysis & Input & Method & Output \\
\midrule
\cellcolor[HTML]{F6EDDC}Questionnaire analysis & Creativity-support ratings adapted from CSI~\cite{cherry2014quantifying,lin2025inkspire,lin2026visuallyrics} and human--AI collaboration ratings~\cite{lawton2023drawing,lin2025inkspire}. & Two-sided Mann--Whitney $U$ tests with Holm correction, applied separately across the six creativity-support items and seven human--AI collaboration items (Section~\ref{sec:study-questionnaires}). & Medians $[Q_1,Q_3]$, rank-biserial correlations, and raw and Holm-adjusted $p$-values for each item. \\
\addlinespace[5pt]
\cellcolor[HTML]{E7EFF7}Interaction sequence analysis & Interaction logs and recordings, organised by participant-confirmed idea boundaries (Section~\ref{sec:interaction-sequence-analysis}). &
\designintentcode{}\newline
Code expressed intentions in each canvas submission.\par\smallskip
\followupcode{}\newline
Code the response to AI Interprets.\par\smallskip
\agencycode{}\newline
Code the full idea sequence.\par\smallskip
Codebooks: Appendix~\ref{sec:interaction-codebooks}.
Analysis details: Section~\ref{sec:interaction-sequence-analysis}. &
\textbf{Design Intent counts:} within each condition, count participants expressing a category at least once across their submissions; each participant contributes at most one count per category.\par\smallskip
\textbf{Human Follow-up and Creative Agency distributions:} identify each participant\textquotesingle s most frequent action or pattern, retain ties as separate classifications, and report participant counts and percentages within each condition (Section~\ref{sec:coding-aggregation}). \\
\addlinespace[5pt]
\cellcolor[HTML]{F1E7EB}Directed qualitative content analysis & Retrospective interview transcripts. & Directed coding~\cite{hsieh2005three} guided by creativity-support~\cite{cherry2014quantifying,lin2025inkspire,lin2026visuallyrics} and human--AI collaboration~\cite{lawton2023drawing,lin2025inkspire} questionnaire dimensions; comparisons within and across conditions (Section~\ref{sec:interview-analysis}). & Coded accounts explaining participants' experiences in the two conditions. \\
\bottomrule
\end{tabularx}
\endgroup
\end{table*}

\subsubsection{Questionnaires}
\label{sec:study-questionnaires}
Participants completed post-task questionnaires measuring perceived creativity support and human--AI collaboration using seven-point Likert scales.
Informed by previous related systems ~\cite{lin2025inkspire, lin2026visuallyrics}, we use the six items from the Creativity Support Index (CSI) for our study.
Its six items cover \textit{exploration}, \textit{inspiration}, \textit{immersion}, \textit{expressiveness},
\textit{enjoyment}, and \textit{results worth effort}.
We report these as item-level scores. 
The human--AI collaboration questionnaire similarly follows items from prior work~\cite{lawton2023drawing,lin2025inkspire}, covering \textit{communication},
\textit{alignment}, \textit{controllability}, \textit{harmony}, \textit{partnership}, \textit{attribution}, and \textit{ownership}.
Both questionnaires used anchors from 1 (Strongly Disagree) to 7 (Strongly Agree); the exact items are reported with the corresponding Results figures.
Questionnaire responses were analysed at the participant level. Because not all
variables satisfied normality assumptions, we used two-sided Mann--Whitney
$U$ tests to compare the two conditions. Descriptive statistics are reported as
medians ($Md$) and interquartile ranges (IQR), with rank-biserial correlation
($\delta$) reported as the effect size.
To account for multiple comparisons, we applied Holm's step-down correction
separately to two families of tests: the six creativity-support items and
the seven human--AI collaboration items. We used an adjusted significance
threshold of $p_{\mathrm{Holm}}<.05$. Tables report both unadjusted and
Holm-adjusted $p$-values, medians with first and third quartiles
($Md\,[Q_1,Q_3]$), and signed rank-biserial correlations; positive values
indicate higher ratings in \ts{}. Figures show the full rating distributions.

To organise the Results narrative, we considered the quantitative and qualitative findings together, including questionnaire responses, interview accounts, and coded interactions. We report all questionnaire items and integrate relevant qualitative evidence to characterise participants' experiences and interaction processes.

\subsubsection{Human--AI Interaction Sequences}
\label{sec:interaction-sequence-analysis}
\begin{figure*}[t]
\centering
\includegraphics[width=\linewidth]{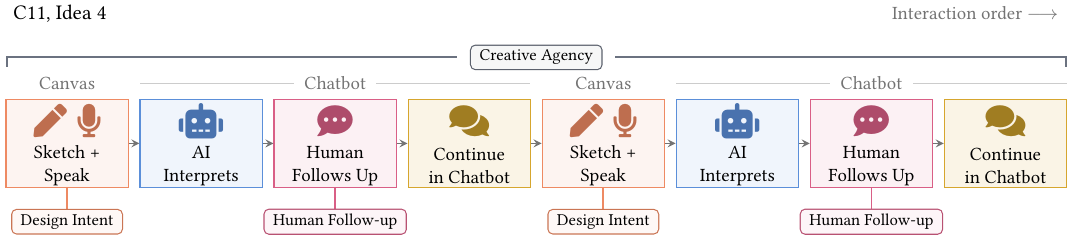}
\caption{Coding scopes within C11's fourth idea. The sequence preserves
the order of two canvas submissions and their subsequent chatbot activity;
consecutive later exchanges are grouped as Continue in Chatbot.
The guiding questions are: Design Intent,
\textbf{what design intentions are expressed?} For this coding, \bl{}
uses only the sketch (including canvas annotations), whereas \ts{} uses
the sketch and concurrent speech together.
Human Follow-up, \textbf{how does the participant respond to
AI Interprets?} Creative Agency, \textbf{how do human and AI
contributions shape the full idea?} The latter two dimensions code the
post-interpretation follow-up and the full idea-level sequence,
respectively, in both conditions.}
\label{fig:code-case}
\Description{Two repeated cycles run left to right: Sketch plus Speak,
AI Interprets, Human Follows Up, and Continue in Chatbot. Orange Design
Intent labels attach to the two canvas submissions; pink Human Follow-up
labels attach to the responses after AI interpretation. A grey Creative
Agency bracket spans the full sequence. All event boxes and coding labels
are rectangular. Guiding questions and the
condition-specific evidence sources are explained in the caption.}
\end{figure*}

The prototype automatically logged interaction events in chronological order,
including sketching actions, concurrent speech, prompts, AI interpretations,
generated text and images, imported results, and system events. After the
task, participants reviewed their interaction history with the researcher
and confirmed the boundaries between successive design ideas. These
participant-confirmed idea instances formed the basis of sequence coding.
We applied three complementary coding dimensions across both conditions:
Design Intent examined what participants expressed about their designs;
Human Follow-up examined their responses to AI interpretations; and
Creative Agency examined how human and AI contributions shaped each idea
(Figure~\ref{fig:code-case}).

\paragraph{\designintentcode{}}
\label{sec:design-intent-coding}
Following prior work~\cite{suwa1998macroscopic, shi2026drawing, gero2014function, desmet2007framework, crilly2004seeing}, we adopted the design-intent coding
scheme to examine what participants expressed in each canvas submission within
participant-confirmed ideas, applying the relevant categories to the scope of
the short toaster-design task. We coded sketches and annotations in \bl{}, and
sketches, annotations, and concurrent speech in \ts{}. Multiple categories
could apply to a submission, with evidence recorded as canvas, speech, or both
(Appendix~\ref{sec:codebook-design-intent}).

\paragraph{\followupcode{}}
Drawing on prior work on human--AI co-creation~\cite{davis2026human}, we
examined how participants responded to AI Interprets. The coding unit was
the participant's follow-up, read together with the preceding AI
interpretation (Appendix~\ref{sec:codebook-followup}).

\paragraph{\agencycode{}}
Drawing on prior work on human--AI collaboration~\cite{shi2026taxonomy,li2026ai},
we examined how human and AI contributions shaped each participant-confirmed
idea. We read the full sequence of canvas activity, AI interpretations,
participant responses, and generated outputs, assigning one pattern to each
idea (Appendix~\ref{sec:codebook-agency}).

\paragraph{Participant-level aggregation and comparison}
\label{sec:coding-aggregation}
For Design Intent, we counted participants who expressed each category at
least once across their submissions. Each participant could contribute to
multiple categories.

For Human Follow-up and Creative Agency, we classified each participant by
the most frequent action across their coded follow-ups or pattern across
their coded ideas. Classification considered all five Follow-up categories
and all four Agency patterns, with equally frequent categories recorded as
ties. We reported these classifications descriptively as counts and
percentages of the twelve participants in each condition.

Image-generation usage was calculated as the number of image-generation
events divided by all AI-use events within each condition.

\subsubsection{Interviews and Directed Qualitative Content Analysis}
\label{sec:interview-analysis}
We conducted semi-structured retrospective interviews with all participants.
The interviews examined how participants communicated ideas, responded to AI
interpretations, incorporated generated outputs, and experienced the assigned
condition. Interviews were audio-recorded, transcribed using
WhisperX~\cite{bain2023whisperx}, and manually checked by the first author
before analysis.

Interview transcripts were analysed using directed qualitative content
analysis~\cite{hsieh2005three}, guided by the creativity-support and
human--AI collaboration questionnaire dimensions. The analysis examined how
participants expressed design intent, experienced communication within
sketching, used AI visualisations, and responded to AI understanding and
misunderstanding. We refined the codes through examination of the transcripts
and compared participants' accounts within and across conditions to explain
how they experienced the two forms of grounding during design ideation.

We counted each participant once per interview theme and compared theme
coverage across conditions. Participants could contribute to multiple themes.

\paragraph{Coding procedure and inter-annotator agreement}
For each of the three interaction coding schemes and the interview analysis,
two researchers iteratively developed a codebook and then independently
coded the data using the corresponding codebook. The three interaction
codebooks are provided in Appendix~\ref{sec:interaction-codebooks}. Inter-annotator agreement
(IAA) was calculated using Cohen's $\kappa$: .79 for Design Intent,
.73 for Human Follow-up, .77 for Creative Agency, and .71 for the
interview analysis.

\section{Results}
We organise the results around the two central outcomes of the study:
designers' perceived creativity support and their experience of human--AI
collaboration. For each outcome, we first report the questionnaire results and
then use interaction sequence analysis and interview findings to contextualise
the observed patterns and describe participants' experiences.


\subsection{Creativity Support: Natural Expression of Design Intent and Efficient Visualisation}

\begin{figure}[htbp]
\centering
\includegraphics[width=\linewidth]{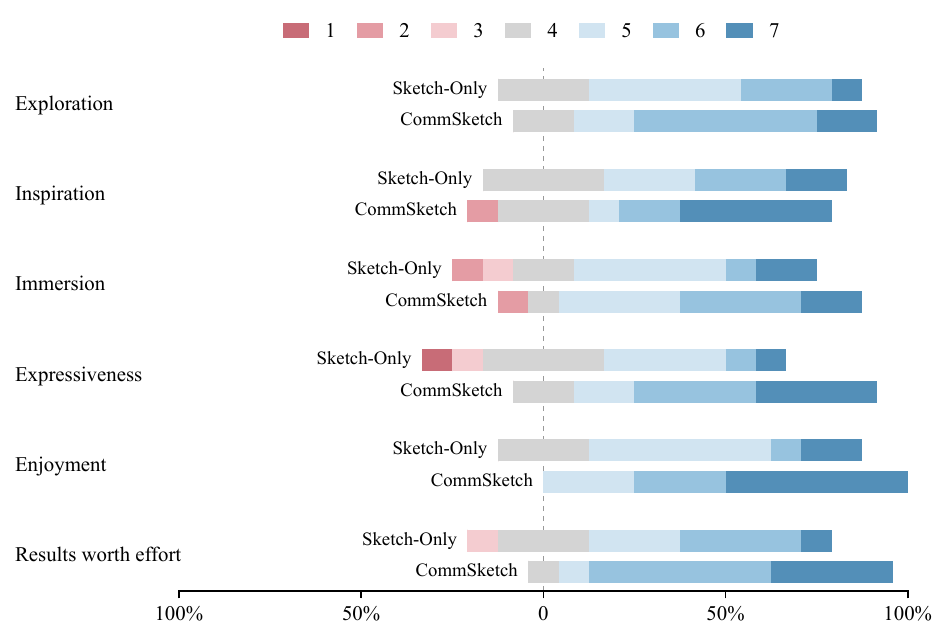}
\caption{Distribution of creativity support ratings using adapted items~\cite{cherry2014quantifying,lin2025inkspire,lin2026visuallyrics} ($n=12$ per condition). Colours represent ratings from 1 (Strongly Disagree) to 7 (Strongly Agree). The midpoint rating of 4 is centred on the dashed line; lower ratings extend left and higher ratings extend right. Each bar represents all twelve responses. The bracket and $^{*}$ indicate a between-condition difference with Holm-adjusted $p<.05$.}
\label{fig:csi-hai}
\Description{Paired diverging stacked bars show the seven response categories for each creativity support item in \bl{} and \ts{}.}
\end{figure}

\begin{table*}[htbp]
\centering
\small
\setlength{\tabcolsep}{5pt}
\caption{Comparison of creativity-support ratings ($n=12$ per condition).}
\label{tab:csi-statistics}
\begin{tabular}{@{}lccrrrr@{}}
\toprule
Item & \bl{} & \ts{} & $U$ & $\delta$ & $p_{\mathrm{raw}}$ & $p_{\mathrm{Holm}}$ \\
 & Median $[Q_1,Q_3]$ & Median $[Q_1,Q_3]$ & & & & \\
\midrule
Exploration & 5.0 [4.75, 6.00] & 6.0 [5.00, 6.00] & 50.0 & 0.31 & $.194$ & $.582$ \\
Inspiration & 5.0 [4.00, 6.00] & 6.0 [4.00, 7.00] & 60.5 & 0.16 & $.512$ & $.619$ \\
Immersion & 5.0 [4.00, 5.25] & 5.5 [5.00, 6.00] & 54.5 & 0.24 & $.309$ & $.619$ \\
Expressiveness & 4.5 [4.00, 5.00] & 6.0 [5.00, 7.00] & 32.0 & 0.56 & $\underline{.019}^{*}$ & $.097$ \\
Enjoyment & 5.0 [4.75, 5.25] & 6.5 [5.75, 7.00] & 31.5 & 0.56 & $\underline{.015}^{*}$ & $.091$ \\
Results worth effort & 5.0 [4.00, 6.00] & 6.0 [6.00, 7.00] & 36.0 & 0.50 & $\underline{.032}^{*}$ & $.127$ \\
\bottomrule
\end{tabular}
\par\smallskip
\begin{minipage}{\linewidth}
\footnotesize
\textit{Note.} Values are medians with first and third quartiles, $Md\,[Q_1,Q_3]$. Two-sided Mann--Whitney $U$ tests use the \bl{} group for $U$; positive rank-biserial correlations ($\delta$) indicate higher ratings in \ts{}. Holm correction is applied across all 6 items within this questionnaire. Underlining marks significant raw $p$-values; bold marks significant Holm-adjusted $p$-values. Stars apply independently to each column: $^{*}p<.05$, $^{**}p<.01$, $^{***}p<.001$.
\end{minipage}
\end{table*}

Expressiveness, Enjoyment, and Results Worth Effort had higher median ratings
in \ts{} and the largest observed effect sizes among the six adapted items
($\delta=.56$, $.56$, and $.50$, respectively; Table~\ref{tab:csi-statistics}).
Their raw $p$-values were .019, .015, and .032; none of the six comparisons
reached significance after Holm correction. We focus on these three dimensions
to interpret their observed rating differences alongside interview and
interaction evidence about participants' creative experiences
(Figure~\ref{fig:csi-hai}).


\subsubsection{\textbf{Expressiveness: From Physical Form to Conceptual Intent}}
\label{sec:expressiveness}

\begin{figure}[t]
\centering
\includegraphics[width=0.6\linewidth]{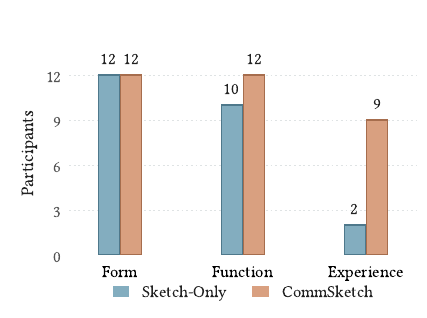}
\caption{Participant-level counts of three Design Intent categories:
\textbf{Form} (physical appearance), \textbf{Function} (features and
intended operation), and \textbf{Experience} (intended use qualities,
sensory outcomes, and affective qualities). Qualitative coding considered canvas
content in \bl{} and canvas content with concurrent speech in \ts{}.
Bars show the number of participants expressing each category at least
once ($n=12$ per condition). Participants could express multiple categories.}
\label{fig:design-intent-counts}
\Description{Grouped bars show the number of participants expressing each
design intention category. \bl{}, in blue: Form 12, Function 10,
Experience 2. \ts{}, in orange: Form 12, Function 12, Experience 9.
Each condition contains 12 participants.}
\end{figure}

Our qualitative coding of \designintentcode{} examined the expression of \textit{Form},
\textit{Function}, and \textit{Experience} across both conditions
(Section~\ref{sec:design-intent-coding}). Across the available design submissions during the 30-minute session, all twelve participants in each condition expressed form-related intentions. Function-related intentions were expressed by twelve \ts{}
participants and ten \bl{} participants. Experience-related intentions were
expressed by nine \ts{} participants, compared with two \bl{} participants.
Each participant was counted once per category, considering canvas content
in \bl{} and canvas content together with concurrent speech in \ts{}
(Figure~\ref{fig:design-intent-counts}).

Both conditions therefore supported the expression of physical form and
intended function. The clearest descriptive difference concerned
experience-related intentions, including portability, ease of use, safety,
and humour. Concurrent speech provided an additional channel for articulating
these qualities alongside the sketch.

\begin{figure}[t]
\centering
\begingroup
\definecolor{portableS}{HTML}{268BD2}
\definecolor{portableC}{HTML}{E8871E}
\captionsetup[subfigure]{font=small,justification=centering,skip=2pt}
\begin{subfigure}[t]{.495\linewidth}
\centering
\begin{minipage}[b][1.35cm][b]{\linewidth}
\centering
\begin{tikzpicture}
\node[draw=portableS,fill=portableS!5,rounded corners=2pt,
inner sep=3pt,align=left,font=\footnotesize]
{\textbf{Canvas annotation}\\\dquote{portable toaster}};
\end{tikzpicture}
\end{minipage}
\par\vspace{0pt}
\includegraphics[width=\linewidth]{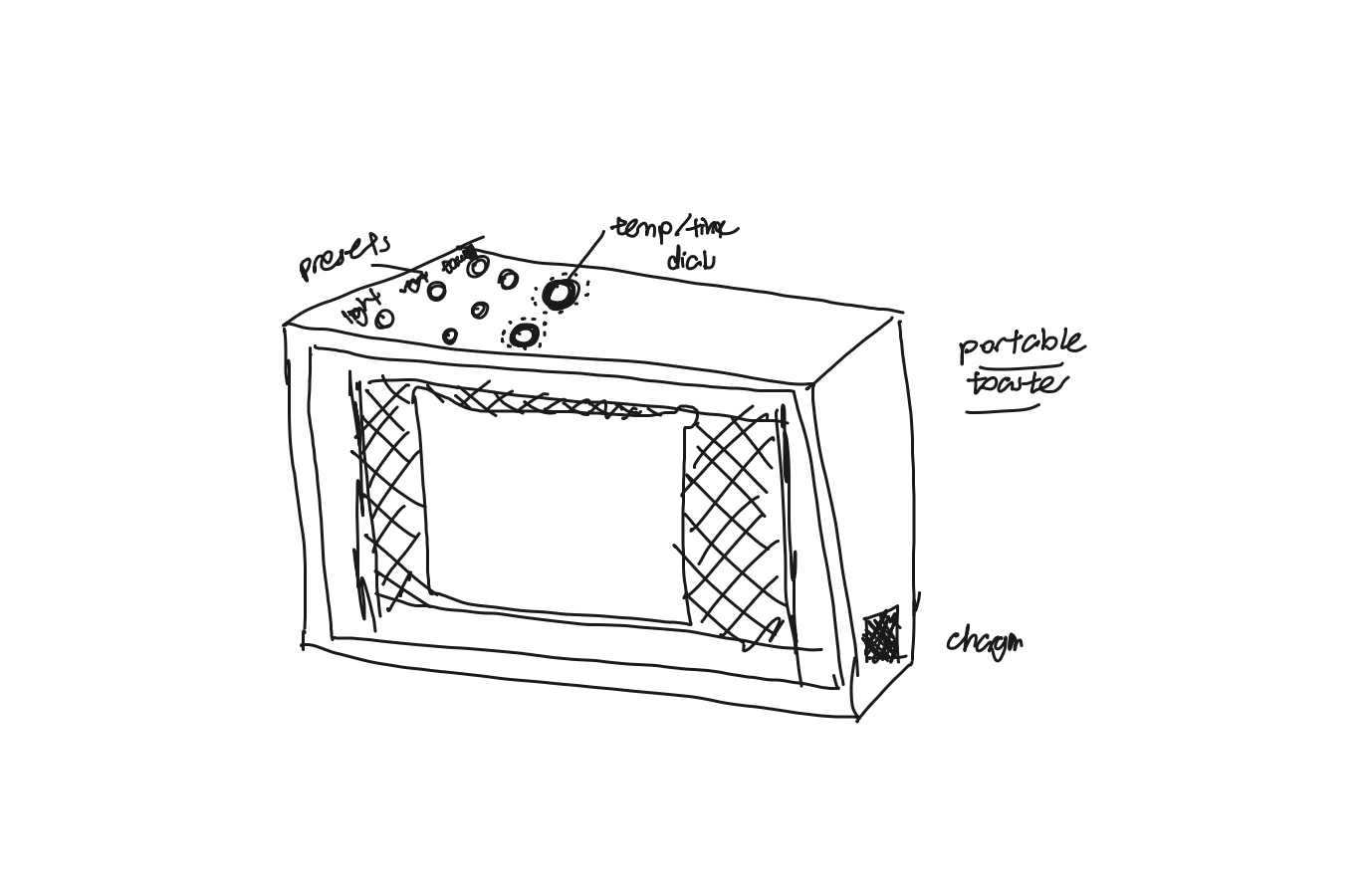}
\par\vspace{-6mm}
\caption{S4: \bl{}}
\end{subfigure}\hfill
\begin{subfigure}[t]{.495\linewidth}
\centering
\begin{minipage}[b][1.35cm][b]{\linewidth}
\centering
\begin{tikzpicture}
\node[draw=portableC,fill=portableC!5,rounded corners=2pt,
inner sep=3pt,text width=.9\linewidth,align=left,font=\footnotesize]
{\textbf{Concurrent speech}\\
\dquote{it would use batteries and a handle.}\ $\ldots$\
\dquote{So that It could be brought around everywhere.}};
\end{tikzpicture}
\end{minipage}
\par\vspace{0pt}
\includegraphics[width=\linewidth]{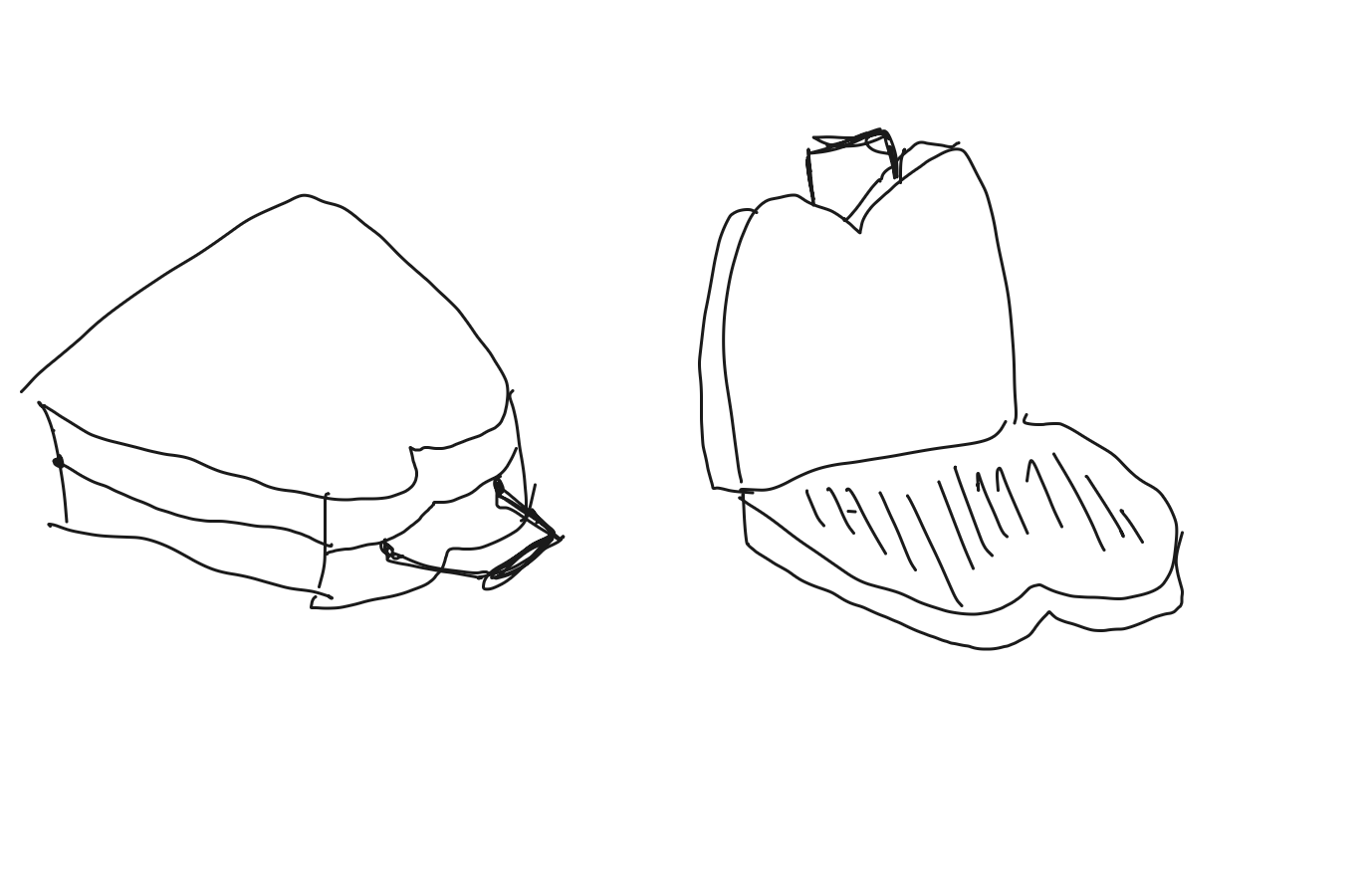}
\par\vspace{-6mm}
\caption{C5: \ts{}}
\end{subfigure}
\endgroup
\caption{Expressing portability in the first canvases of S4's third and
C5's sixth ideas. S4 states portability in an annotation; C5's concurrent
speech relates carrying and on-the-go use to the proposed design features.
Excerpts reproduce canvas text (blue) and speech (orange); the ellipsis
marks omitted speech.}
\label{fig:portable-expression}
\Description{Two complete initial canvases side by side. S4 draws a box-like
toaster with labelled controls and the words portable toaster. C5 draws a
bread-shaped hinged toaster in open and closed views. A speech excerpt
above C5's canvas mentions batteries, a handle, and carrying it around.}
\end{figure}
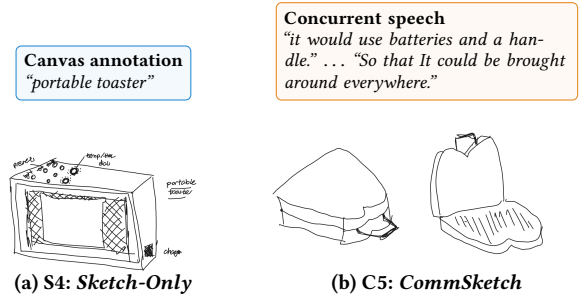

Figure~\ref{fig:portable-expression} illustrates this difference through
participants' expressions of portability. In \bl{}, S4 wrote
\dquote{portable toaster} beside the sketch, directly naming the intended
quality. In \ts{}, C5 described a smaller, lunchbox-like toaster with battery
power and a handle, explaining while sketching how it could be carried and
used on the go. Speech therefore connected the intended experience of
portability to concrete aspects of the design's form and function. Similarly,
while sketching a basket-shaped toaster, C8 explained that
\dquote{people can carry it out}, linking an otherwise ambiguous visual form
to its intended use. These examples show how concurrent speech could express
not only what a design looked like, but what its features were intended to
enable.

Nine of twelve \ts{} participants described speech as helping them communicate design meanings,
reasoning, or contextual information that were difficult to convey through
visual marks alone. For example, C5 explained that
\dquote{a lever might not look like a lever in the sketch}, but
\dquote{if we talk and say that it's a lever, then the AI would understand}.
C10 similarly noted that without speech,
\dquote{the AI wouldn't have the context of what I'm trying to sketch or
create}. Speech therefore allowed participants to assign intended meanings to
rough or ambiguous visual marks without first making every element visually
self-explanatory.

Four \ts{} participants specifically described speaking
their thoughts while drawing as natural or valued the synchronisation between
the two modalities. C7 noted that
\dquote{the sketch and explanation are synchronized} and that
\dquote{Everything happens at the same time}. Participants could also
articulate functional details and reasoning as they emerged. C11, for
example, described explaining how a knob should turn or how a
rice-cooker-like mechanism should work, noting that
\dquote{Being able to verbalize my thoughts and explain my reasoning while
ideating was very helpful}. C11 added that explaining these details only
after drawing could require reconstructing the earlier thought process:
\dquote{I might not even remember all the details I was thinking about during
sketching}.

Together, these findings show that more \ts{} participants expressed experience-related intentions during the session. The examples and interview accounts suggest that concurrent speech helped participants articulate intended qualities, functional meanings, and reasoning alongside their sketches.


\subsubsection{\textbf{Enjoyment: From Separate Explanation to Embedded Communication}}

Participants valued keeping communication with AI within their ongoing
sketching activity.

Eight of twelve \ts{} participants
described speaking while sketching as allowing them to express ideas naturally
or synchronously, or as reducing the need to type, provide separate
explanations, or recall details afterward. Four of these participants
specifically emphasised the naturalness or
synchronisation of sketching and explanation. C9 explained that because
speech occurred during sketching, \dquote{that information comes out
naturally}. C9 also distinguished this verbalisation from explicit prompting:
\dquote{we're expressing our thoughts} and
\dquote{We don't necessarily think of them as instructions}. C7 similarly
described sketching and explanation as a continuous interaction in which
\dquote{the sketch and explanation are synchronized... everything happens
at the same time}.
Five \ts{} participants explicitly expressed a
liking for or preference for speech-based interaction. C12 stated,
\dquote{Personally, I find it easier to just say what I'm thinking rather
than type it all out}, while C6 remarked,
\dquote{Typing would be a hindrance. Talking was the best option}. C11
similarly stated, \dquote{I really liked being able to talk while drawing}.

Three \bl{} participants described challenges
or strategies for coordinating sketching with separate explanations
and chatbot interaction. S7 explained that the AI initially did not
understand their sketches, but \dquote{as I added prompts, it understood}.
S8 described a workflow in which they would first explain an idea and then
upload the sketch so that the AI already knew \dquote{what to expect}. S3
more broadly felt that a linear chatbot did not fit sketching and ideation
well because activities such as learning, sketching, and generating concepts
were \dquote{mixed together inside one conversation}.

Participants valued being able to communicate within the flow of sketching,
rather than reconstructing their ideas afterward. Their accounts describe
verbalising ideas as they formed, with less need to shift into a separate
mode of explanation or explicit prompting.


\subsubsection{\textbf{Results Worth Effort: From Rough Sketches to Usable Renderings}}

AI-generated images helped participants visualise and develop their designs.
All twelve \ts{} participants and eleven of twelve \bl{} participants
described a useful role for AI visualisation. In both conditions, AI use was
largely centred on generating images: among AI-use events, image generation
accounted for 77.4\% in the \bl{} condition and 86.3\% in the \ts{} condition.
Four \ts{} participants and three \bl{} participants specifically highlighted
faster visualisation or visual iteration.

Obtaining useful AI-generated outputs could nevertheless require additional
effort, including refining sketches, providing further explanations, and
correcting or clarifying mismatched outputs. The two conditions differed in
how this effort was incorporated into the design process.

In \bl{}, additional effort could be directed toward making the sketch or
accompanying prompt sufficiently explicit for the AI. Five \bl{}
participants specifically felt that the AI relied more on written prompts or
annotations than on the sketch itself. S6, for example, preferred to improve
the sketch rather than write a longer prompt, directing this effort back into
the visual artefact. S8 similarly described providing explanations before
showing a sketch so that the AI knew \dquote{what to expect}.

In \ts{}, speech provided an additional way to carry design intent without
requiring the sketch itself to become fully explicit. Six \ts{} participants
specifically described turning rough or simple sketches into usable design
representations. C5 explained that although it was \dquote{very hard to draw
quickly and generate a lot of ideas}, they could \dquote{sketch something
roughly} and obtain a \dquote{pretty good-looking mock-up} that could be shown
to others. C8 similarly noted that even \dquote{very simple sketches} could be
rendered into something \dquote{easier to understand and easier to visualize}.
As shown in Section~\ref{sec:expressiveness}, speech helped participants
express intent beyond what was visible in the sketch itself. C5 noted that
talking helped the AI understand the \dquote{big picture}, while C9 explained
that, rather than having to specify design details deliberately,
\dquote{that information comes out naturally} while sketching.

These accounts suggest that speaking while sketching made AI-generated images
feel more worth the effort by changing how that effort was incorporated into
the design process. Rather than making a rough sketch sufficiently
self-explanatory before rendering, participants could articulate intent
alongside it and use AI to turn the developing idea into a usable visual
representation.


\subsection{Human--AI Collaboration: Establishing a Shared Design Premise for Alignment and Co-evolution}
\begin{figure}[htbp]
\centering
\includegraphics[width=\linewidth]{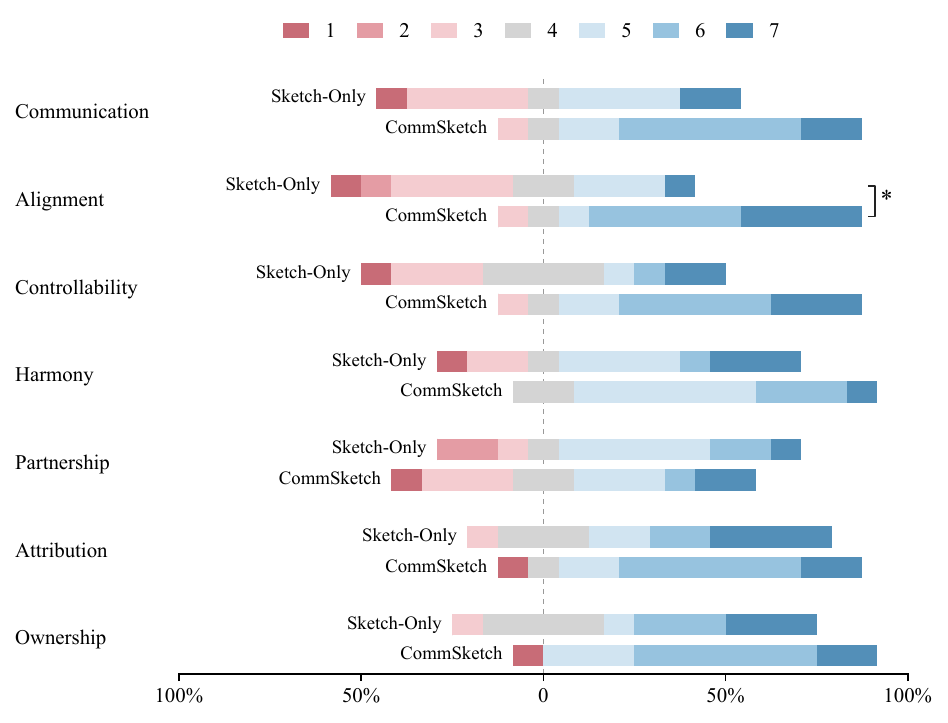}
\caption{Distribution of human--AI collaboration ratings using adapted items~\cite{lawton2023drawing,lin2025inkspire} ($n=12$ per condition). Colours represent ratings from 1 (Strongly Disagree) to 7 (Strongly Agree). The midpoint rating of 4 is centred on the dashed line; lower ratings extend left and higher ratings extend right. Each bar represents all twelve responses. The bracket and $^{*}$ indicate a between-condition difference with Holm-adjusted $p<.05$.}
\label{fig:hai}
\Description{Paired diverging stacked bars show the seven response categories for each human--AI collaboration item in \bl{} and \ts{}. A bracket and one star mark the Alignment comparison, with Holm-adjusted p = .028.}
\end{figure}
\begin{table*}[htbp]
\centering
\small
\setlength{\tabcolsep}{5pt}
\caption{Comparison of human--AI collaboration ratings ($n=12$ per condition).}
\label{tab:hcai-statistics}
\begin{tabular}{@{}lccrrrr@{}}
\toprule
Item & \bl{} & \ts{} & $U$ & $\delta$ & $p_{\mathrm{raw}}$ & $p_{\mathrm{Holm}}$ \\
 & Median $[Q_1,Q_3]$ & Median $[Q_1,Q_3]$ & & & & \\
\midrule
Communication & 4.5 [3.00, 5.00] & 6.0 [5.00, 6.00] & 37.5 & 0.48 & $\underline{.045}^{*}$ & $.268$ \\
Alignment & 3.5 [3.00, 5.00] & 6.0 [5.75, 7.00] & 22.5 & 0.69 & $\underline{.004}^{**}$ & $\mathbf{.028}^{*}$ \\
Controllability & 4.0 [3.00, 5.25] & 6.0 [5.00, 6.25] & 38.0 & 0.47 & $\underline{.049}^{*}$ & $.268$ \\
Harmony & 5.0 [3.75, 6.25] & 5.0 [5.00, 6.00] & 65.0 & 0.10 & $.695$ & $1.000$ \\
Partnership & 5.0 [3.75, 5.25] & 4.5 [3.00, 5.25] & 77.0 & -0.07 & $.790$ & $1.000$ \\
Attribution & 5.5 [4.00, 7.00] & 6.0 [5.00, 6.00] & 69.5 & 0.03 & $.905$ & $1.000$ \\
Ownership & 5.5 [4.00, 6.25] & 6.0 [5.00, 6.00] & 61.5 & 0.15 & $.549$ & $1.000$ \\
\bottomrule
\end{tabular}
\par\smallskip
\begin{minipage}{\linewidth}
\footnotesize
\textit{Note.} Values are medians with first and third quartiles, $Md\,[Q_1,Q_3]$. Two-sided Mann--Whitney $U$ tests use the \bl{} group for $U$; positive rank-biserial correlations ($\delta$) indicate higher ratings in \ts{}. Holm correction is applied across all 7 items within this questionnaire. Underlining marks significant raw $p$-values; bold marks significant Holm-adjusted $p$-values. Stars apply independently to each column: $^{*}p<.05$, $^{**}p<.01$, $^{***}p<.001$.
\end{minipage}
\end{table*}

Alignment showed the largest observed effect and remained significant after
Holm correction ($\delta=.69$, $p_{\mathrm{Holm}}=.028$;
Table~\ref{tab:hcai-statistics}). Communication and Controllability had the
next-largest effects ($\delta=.48$ and $.47$), with higher median ratings
in \ts{}, but neither reached significance after correction
(both $p_{\mathrm{Holm}}=.268$). No other item reached adjusted significance.
We examine these three dimensions together because their rating distributions
(Figure~\ref{fig:hai}) and related interview and interaction evidence illuminate
how participants communicated intent, established alignment, and directed AI
contributions.


\subsubsection{\textbf{Communication: From Ambiguous Sketches to a Shared Design Premise}}

Whereas Expressiveness (Section~\ref{sec:expressiveness}) concerned what
participants could communicate through sketching and speech, Communication
concerned whether this information established a workable design premise for
the AI. In \bl{}, AI Interprets was based on the sketch alone; in \ts{}, it
was based on both the sketch and concurrent speech.

Six \bl{} participants described communication
difficulties arising from unclear sketches, missing visual cues, spatial
interpretation, or ambiguous markings. In \bl{}, these ambiguities sometimes
affected the AI's understanding of the basic design premise. A triangular
toaster was interpreted as a tent, a disk-player-shaped toaster as a game or
puzzle, and a bus-shaped toaster as an actual bus. S10 attributed one
misunderstanding to a missing knob, while S6 noted that a conventional
box-shaped toaster was likely to be recognised whereas a \dquote{weird shape}
was not.

When this premise was missed, participants had to supply it explicitly in
their subsequent follow-up. Examples included \dquote{This is a triangular
toaster} (S7), \dquote{It's supposed to be a toaster which looks like a yellow
bus} (S10), and simply \dquote{Toaster} (S6). These responses did more than
refine local details: they re-established what the design was before the
interaction could proceed.

This need was also reflected in participants' design recommendations. Three
\bl{} participants and one \ts{} participant suggested that the AI should
clarify the intended object, concept, or functionality before proceeding when
the input was ambiguous. S7, for example, argued that the AI
\dquote{shouldn't assume things too early}.

These findings suggest that concurrent speech could make more of the intended
design premise available before AI Interprets was produced. The next question,
then, is whether establishing this premise earlier changed what participants
needed to do after seeing the AI's interpretation.


\subsubsection{\textbf{Alignment: From Repairing AI Interpretations to Acting on Them}}

\begin{figure*}[htbp]
\centering
\includegraphics[width=0.8\textwidth]{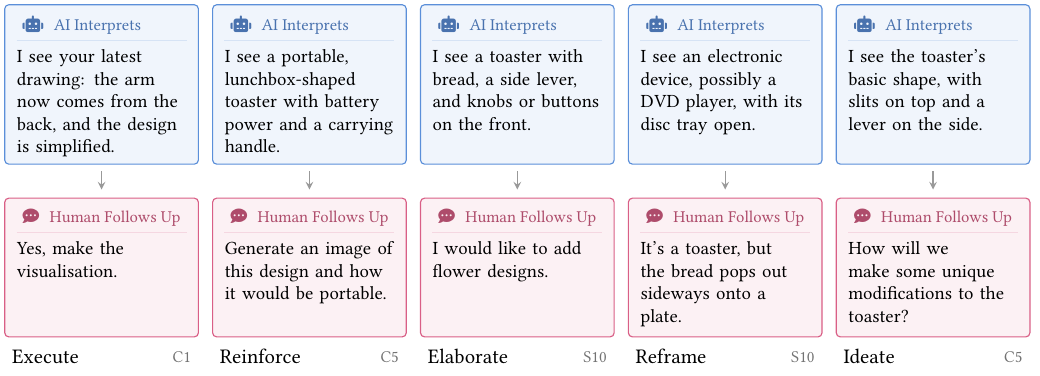}
\caption{Examples of five Human Follow-up codes after AI Interprets (Appendix~\ref{sec:codebook-followup}):
Execute, Reinforce, Elaborate, Reframe, and Ideate. Each column pairs an AI
interpretation with the participant's subsequent response. Dialogues are
condensed for readability. Blue messages represent AI Interprets and pink
messages represent Human Follows Up, matching the qualitative interaction coding figure.}
\label{fig:response_codes}
\Description{Five columns pair AI interpretations with responses by C1,
C5, S10, S10, and C5, respectively illustrating execution, reinforcement,
elaboration, reframing, and ideation.}
\end{figure*}

\begin{figure}[htbp]
\centering
\includegraphics[width=\linewidth]{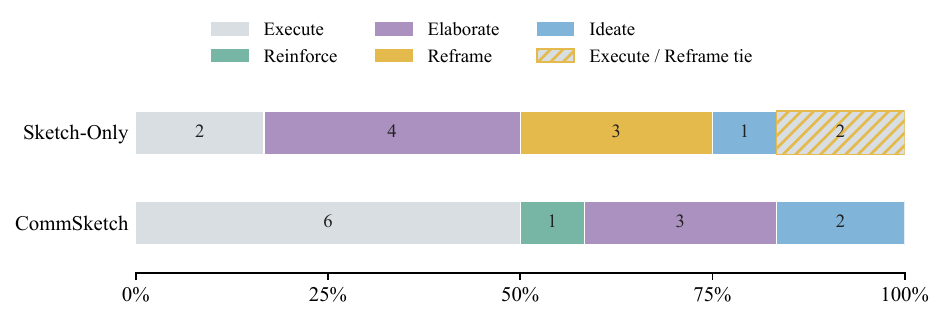}
\caption{Distribution of participants by their most frequent Human Follow-up action
(Appendix~\ref{sec:codebook-followup} and $n=12$ per condition). Grey indicates Execute, teal Reinforce, purple Elaborate,
gold Reframe, and blue Ideate. Stripes indicate equally frequent Execute
and Reframe actions. Segment labels show participant counts; bar widths show percentages.}
\label{fig:interaction-code}
\Description{Stacked bars compare participants' most frequent follow-up actions. Execute is most frequent for six \ts{} and two \bl{} participants. Reframe is uniquely most frequent for three \bl{} participants and none in \ts{}.}
\end{figure}

We therefore examined participants' Human Follows Up after AI Interprets.
Our qualitative coding of \followupcode{} identified five response types: Execute, Reinforce,
Elaborate, Reframe, and Ideate. Figure~\ref{fig:response_codes} illustrates
each response type by pairing an AI interpretation with the participant's
subsequent response. The participant distributions showed contrasting
patterns of execution and reframing (Figure~\ref{fig:interaction-code}).


Reframe was the most frequent action for 25.0\% of participants in \bl{}
and none in \ts{}; a further 16.7\% in \bl{} had equally frequent Execute
and Reframe actions (Figure~\ref{fig:interaction-code}). Execute was the most
frequent action for 50.0\% of participants in \ts{}, compared with 16.7\%
in \bl{}.
In \bl{}, Reframe follow-ups often
restated the object, supplied missing premise information, or redirected an
incorrect interpretation before participants could continue.

Once an interpretation was usable, participants could instead act on or
develop it. Examples included \dquote{Generate this one} and \dquote{Yes, make the visualisation}
(C1). Participants also used AI Interprets as starting points for further
ideation. C10, for example, asked \dquote{Could you recommend me any areas of
improvement?} and later requested more specific advice for making the design
look metallic and three-dimensional.

The predominance of Execute rather than Reframe among more \ts{} participants
suggests that acting on an interpretation was more characteristic of their
responses than repairing it. This provides behavioural context for
the higher perceived Alignment in \ts{} and shifts attention from correcting
what the AI understood to deciding what to do with its contribution.

\subsubsection{\textbf{Controllability: From Directing Execution to Guiding Exploration}}

Participants reported higher median Controllability in \ts{} than in
\bl{} ($Md_{TS}=6.0$ vs.\ $Md_{BL}=4.0$; $U=38.0$, $p=.049$,
$\delta=0.47$; $p_{\mathrm{Holm}}=.268$), although the difference was not
significant after Holm correction. This item concerns feeling in control of
the AI's behaviour. The interviews and interaction examples illustrate how
participants directed what the AI should do and selected which contributions
to incorporate, from requesting the execution of specified designs to
inviting suggestions for further exploration.

Seven \ts{} participants and five \bl{} participants described correcting AI
contributions or selectively retaining those that suited their intentions.
Directing execution could require repeated intervention. S6, for example, added a
hinge and anticipated needing \dquote{more time and more iterations} to obtain
the intended result, while S11 described discrepancies that still required
adjustment despite detailed prompts. These accounts describe the effort
required to steer generation toward specified features.

Selective acceptance also appeared in both conditions. S2 explained that an
alternative interpretation could be retained if it was \dquote{actually
better}, but otherwise \dquote{I would correct it}. C10 similarly described
using suitable AI suggestions while retaining decision authority:
\dquote{If I can tell that something is wrong, then I'll make the correct
changes myself}.

C8 also described modifying generated images across iterations and
experimenting with materials and colours to build on an idea. C7 described
inviting AI input before completing a thought: unlike workflows where
\dquote{When I finish a sketch, I've already finished my thinking}, C7 could
request generation or opinions while the idea was still developing. These
accounts illustrate ways of initiating and guiding further AI contributions.

\begin{figure}[htbp]
\centering
\includegraphics[width=\linewidth]{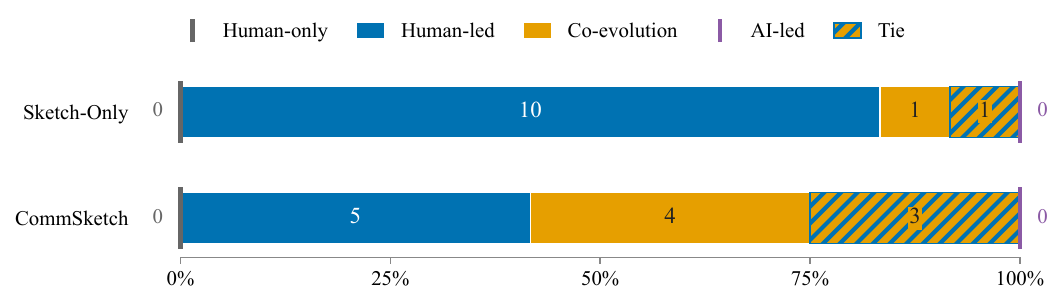}
\caption{Distribution of participants by their most frequent Creative Agency
pattern (Appendix~\ref{sec:codebook-agency} and $n=12$ per condition). Blue indicates Human-led, orange indicates
Co-evolution, and blue-and-orange hatched segments indicate ties between the two.
Human-only and AI-led are shown as grey and purple vertical marks,
respectively, labelled 0 at the bar ends because neither was any participant's
most frequent pattern. Segment labels show participant counts; bar widths
show percentages.}
\label{fig:creative-agency}
\Description{Two stacked horizontal bars compare Sketch-Only and CommSketch.
Human-led accounts for 10 and 5 participants, Co-evolution for 1 and 4,
and ties between the two for 1 and 3, respectively. Human-only and AI-led
have zero participants in both conditions and are shown as vertical marks
labelled 0 at the left and right ends of each bar.}
\end{figure}

The \agencycode{} patterns contextualize these acts of direction and
selection within the broader development of each idea. They describe how
human and AI contributions shaped the idea across the interaction sequence.
Descriptively, Co-evolution was the most frequent pattern for four of twelve
participants (33.3\%) in \ts{}, compared with one of twelve (8.3\%) in \bl{}
(Figure~\ref{fig:creative-agency}). These patterns describe the contributions
to idea development rather than participants' perceived controllability.

\begin{figure*}[t]
\centering
\includegraphics[width=0.8\linewidth]{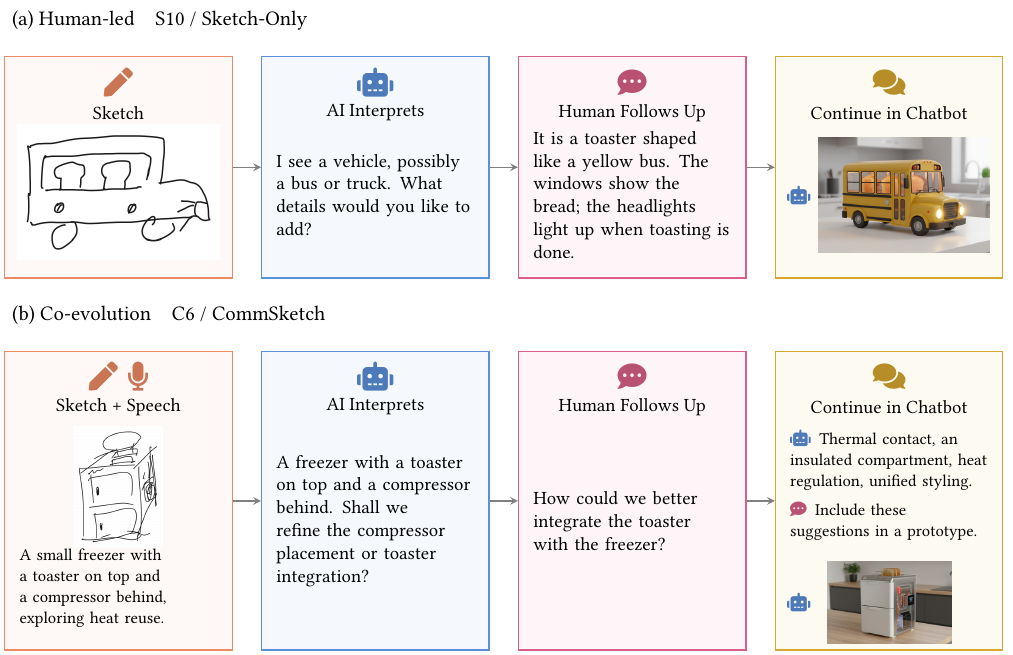}
\caption{
Representative examples of two Creative Agency patterns.
(\textbf{a}) Participant S10 in the \bl{} condition illustrates Human-led
agency by correcting a vehicle interpretation and specifying the toaster's
features before visualization.
(\textbf{b}) Participant C6 in the \ts{} condition illustrates Co-evolution:
after sketching and describing a freezer--toaster concept, C6 asked the AI
how to improve its integration, received suggestions, and requested a
prototype incorporating them. Both rows run left to right through Sketch
(with concurrent speech in \ts{}), AI Interprets, Human Follows Up, and
Continue in Chatbot. The third stage shows the participant's immediate
response; subsequent AI replies, generation requests, and generated images
are grouped in the fourth stage. Excerpts are
condensed. Original sketches and generated images are shown, with only blank
canvas margins removed.
}
\label{fig:trajectory-example}
\Description{Two horizontal rows compare S10's Human-led bus-shaped toaster
and C6's Co-evolution freezer--toaster concept. The top row shows a sketch, AI
interpretation, human clarification, and generated visualization. The bottom
row shows a sketch with speech, AI interpretation, and the participant's
question in Human Follows Up. Continue in Chatbot groups the AI suggestions,
the participant's request to incorporate them, and the generated visualization.}
\end{figure*}

Figure~\ref{fig:trajectory-example} illustrates how participants directed AI
contributions within Human-led and Co-evolution interactions.
In the Human-led example, S10 in \bl{} specified
\dquote{a toaster which looks like a yellow bus} and its features before
visualization, directing the AI to implement an existing concept.
In the Co-evolution example, C6 in \ts{} sketched and described a freezer
with a toaster on top.
After AI Interprets identified the components, C6 asked how the toaster could
be better integrated with the freezer. The AI proposed thermal integration,
an insulated compartment, heat regulation, and a unified appearance. C6 then
requested that the AI \dquote{include that and show me a prototype}.
Here, the participant set the question for exploration and then directed
the incorporation of the AI's suggestions into the next visualization.

Across these examples, participants guided AI behaviour by specifying what
to implement, inviting input on what to explore, and deciding how to use the
resulting contributions. Such direction was visible in both Human-led and
Co-evolution interactions.

\section{Discussion}

Our findings motivate three directions for supporting emerging design intent:
dynamic alignment, broader multimodal expression, and creative participation
as AI takes a greater role in developing ideas.

\subsection{Toward Dynamic Alignment as Design Intent Evolves}

Participants reported higher perceived alignment in \ts{}
($p_{\mathrm{Holm}}=.028$). The follow-up patterns and examples also suggested
that establishing an appropriate interpretation gave subsequent AI
contributions a more useful basis. Recognising a bus-shaped object as a
toaster, for example, allows suggestions to address its intended function.
Prior work showed that concurrent speech can improve generated images'
alignment with design intent~\cite{shi2026drawing}. Our findings complement
this outcome-based evidence by suggesting how speech supports the process
of establishing intent and guiding AI contributions. Together, these
observations motivate understanding alignment as something developed
through interaction, as intentions become available and interpretations
are inspected and corrected.

In our prototype, opening the chatbot supplied the current sketch and
concurrent speech transcript together for a joint interpretation. Speech
could clarify functions and relationships left ambiguous by visual marks,
while the designer could inspect and correct the interpretation before
proceeding. However, this aggregate context was interpreted at user-triggered
checkpoints. The system did not explicitly associate utterances with
particular strokes or track how their relevance changed through erasure,
redrawing, and spoken revision. A reference to a feature might therefore
remain in the transcript after that feature had changed. Supporting AI
understanding in early-stage design requires examining how these evolving
expressions should be related, beyond simply supplying both modalities.

Future work could investigate how speech and canvas edits jointly signal
the introduction, revision, or abandonment of an intention. Erasure alone
need not imply rejection: a designer might redraw the same idea or reconsider
only one of its features. Linking utterances to canvas regions and revisions
could help maintain a more relevant interpretation, while letting designers
inspect and correct which intentions remain active. Building on support
for clarification and correction~\cite{amershi2019guidelines}, fine-grained
studies could examine when AI should update its understanding, ask for
clarification, or leave an ambiguity unresolved. This would test whether
checkpoint-based interpretation can develop into dynamic alignment without
interrupting tentative exploration.

\subsection{Supporting Multimodal Expression of Emerging Design Intent}

More \ts{} participants expressed experience-related intentions during the
session, including portability, ease of use, and safety. Participants also
described articulating functional details and reasoning alongside rough
visual forms. Median Expressiveness ratings were higher, although the
difference was not significant after Holm correction
($p_{\mathrm{Holm}}=.097$). The qualitative evidence provides a starting
point for understanding what speech makes expressible, while the ratings
do not establish a general improvement in perceived expressiveness.
In particular, the examples suggest that designers can communicate how a
design should be used or experienced before its visual form conveys those
intentions clearly.

Concurrent speech made room for these intentions within the act of
sketching. C9 described expressing thoughts without necessarily treating
them as instructions, while C11 valued articulating details before forgetting
them after drawing. These accounts suggest that expression may serve to
explore an emerging possibility as well as communicate an already formed
idea. Interfaces should therefore accommodate incomplete, ambiguous, and
hesitant expression without requiring immediate commitment to a precise
design or command. Our toaster task elicited use qualities that participants
could readily verbalise, however. Speech may help bridge a gap between
knowing what a design should achieve and knowing how to depict it, but
some experiences may remain difficult to put into words.

Spatial and embodied design offer useful contexts for extending this work.
When designing an interactive installation, a designer might demonstrate a
reaching movement, trace a path through a space, or enact the timing of a
bodily response. Put-that-there combined speech and pointing for spatial
reference~\cite{bolt1980put}, while GesPrompt used co-speech gestures to
convey spatial-temporal information to an LLM-based VR
interface~\cite{hu2025gesprompt}. These approaches motivate studying how
speech, sketch, gesture, and bodily action could express aspects of intended
experience that any single modality leaves underspecified. Future studies
should examine which combinations support such intentions across design
tasks, including tentative alternatives, and how designers can express
uncertainty while exploring them.

\subsection{Supporting Human--AI Co-creativity through Speech in Design Ideation}

The interaction examples suggest that supporting co-creativity involves
enabling designers to guide AI participation. C6, for example, invited
integration suggestions and directed their incorporation into the design
(Figure~\ref{fig:trajectory-example}). Such exchanges motivate tools that
let designers shape what AI contributes as their needs change, from
implementing specified features to exploring possibilities and developing
selected suggestions.

C7's account offers a possible explanation for how speech supported AI
participation: the participant could request generation or opinions before
finishing a thought, allowing AI input to enter while the idea was still
developing. Expression and alignment provide additional context for this
account. Concurrent speech made intended uses, experiences, and reasoning
available during sketching, potentially giving AI a basis for suggestions
that participants could develop further. Higher perceived alignment is
consistent with this interpretation. Speech may therefore have created
opportunities for designers to invite and guide AI contributions before
their intentions had settled into a fully specified design.

Existing systems illustrate ways to support these exchanges: Inkspire offers a
sketch-to-design-to-sketch feedback loop for exploring product
concepts~\cite{lin2025inkspire}, while VRCopilot uses intermediate wireframes
to help users guide generated 3D layouts~\cite{zhang2024vrcopilot}.
Both provide opportunities to shape subsequent AI contributions through
continued interaction. This emphasis also aligns with COFI's attention
to communication and the organisation of contributions in
co-creativity~\cite{rezwana2023cofi}. Our findings suggest a complementary
role for speech: making the intentions behind these actions available
while they are still developing. Future systems could let designers indicate
whether they want AI to implement an existing intention, propose alternatives,
or help develop a selected direction, while carrying the relevant speech
and sketch context into subsequent exchanges. Designers could then select
which suggestions to retain and guide their incorporation into revisions.
Evaluating how such exchanges
support exploration and the creativity of resulting designs would extend
our process-level observations of co-evolution toward understanding how
to support co-creativity.
\section{Limitations and Future Work}
\label{sec:limitations}

Our study involved 24 participants, predominantly students with design
coursework or training, alongside participants with professional design
experience. Many reported frequent AI use for design
(Table~\ref{tab:participants}). The small sample may limit the precision
and generalisability of our findings; further studies should include
a wider range of users with varying design expertise and familiarity with
AI to examine how their backgrounds shape
the use and perceived value of speech.

Participants were asked to complete a simple toaster-design task
individually, which may not capture the demands of longer or more complex
design projects. Future work could examine tasks that require balancing
requirements such as portability, capacity, and manufacturing cost, to
understand how designers express and revise these trade-offs through speech
and sketch. Collaborative design sessions could also reveal how speech
supports negotiating a shared design direction. Supporting such sessions
would require distinguishing different speakers, associating their speech
with contributions and editing actions on the shared canvas, and clarifying
when speech is addressed to another designer or to AI.

Our implementation supplied the current sketch and accumulated speech
transcript together, without explicitly linking specific parts of speech to
individual strokes or editing actions. This limits the system's ability to
track which spoken intentions remain relevant as the sketch changes.
Future work could associate speech with specific sketch elements and their
revisions, helping AI track how design intentions evolve through drawing,
erasure, and modification.

Speech was also supplied as transcribed text, which may contain recognition
errors and does not preserve vocal information such as intonation and
emphasis. We did not systematically evaluate the effects of these errors
or information losses on AI interpretation. Future work could explore direct
audio input to investigate whether retaining these vocal cues helps AI
interpret emerging design intentions.

We also did not systematically examine how individual speaking habits
affected interaction or perceived benefits. Designers may differ in their
comfort with verbalising tentative ideas and in how much they choose to
say. Future work could examine how these preferences shape the use of
speech, including whether allowing designers to move between silent
sketching and speaking while sketching supports their preferred ways of
working. This would help identify when speech provides useful context and
when verbalising ideas adds effort to the design process.

Practical adoption also requires integrating speech input into existing
sketching and design platforms. Future development could connect speech
capture to canvas edits and preserve the associated context when designers
revise or transfer sketches for further development. Such integration
should let designers inspect and edit the speech context supplied to AI
within their usual tools. Evaluation in everyday design workflows could
assess whether this reduces the effort of re-explaining ideas across tools
and stages of a project.

\section{Conclusion}
\label{sec:conclusion}

We examined how speaking while sketching steers human--AI design ideation
through a sketch-based AI interface and a study with 24 participants.
Concurrent speech made intended experiences and reasoning available alongside
emerging visual forms. Participants valued communicating within the flow of
sketching and using this context to guide AI visualisations. Speech also
supported perceived alignment, while interaction examples illustrated how
participants could direct AI contributions as ideas developed. Together,
these findings highlight speaking while sketching as a way to communicate
and ground emerging design intent within the activity of ideation. They
motivate AI design tools that support this ongoing expression and allow
designers to shape how human and AI contributions develop together.

\bibliographystyle{ACM-Reference-Format}
\bibliography{main}

\appendix
\onecolumn
\section{Interaction Coding Codebooks}
\label{sec:interaction-codebooks}

The three codebooks address different parts of the interaction: expressed
design intent, the immediate response to AI Interprets, and development of
the full idea. Examples below are condensed descriptions of cases reported
in the Results, rather than additional verbatim quotations.

\subsection{Design Intent}
\label{sec:codebook-design-intent}

\textbf{Guiding question:} What design intentions are expressed during sketching?
The unit is an available canvas submission within a participant-confirmed
idea. Evidence comprises the sketch and canvas annotations in \bl{}, and
the sketch, annotations, and concurrent speech in \ts{}. Categories are
not mutually exclusive. Later chatbot exchanges and AI interpretations are
not used to infer what was expressed in the submission. The categories
follow prior work~\cite{suwa1998macroscopic, shi2026drawing, gero2014function, desmet2007framework, crilly2004seeing}.

\begingroup
\small
\setlength{\tabcolsep}{5pt}
\renewcommand{\arraystretch}{1.15}
\begin{longtable}{@{}p{.15\linewidth}p{.40\linewidth}p{.40\linewidth}@{}}
\caption{Design Intent codebook.}\label{tab:codebook-design-intent}\\
\toprule
Code & Definition and scope & Distinction and example \\
\midrule
\endfirsthead
\toprule Code & Definition and scope & Distinction and example \\
\midrule
\endhead
\bottomrule
\endfoot
\textbf{Form} &
Physical appearance, including shape, size, material, colour, and the
arrangement of components. &
Appearance alone does not establish an intended function or experience.
Example: a basket-shaped body or a lunchbox-like form with a handle. \\
\addlinespace[6pt]
\textbf{Function} &
Features and intended operation, including what components do and how
controls or mechanisms work. &
Code the operation rather than automatically inferring its benefit to a
user. Example: C11 explains how a knob turns or how a rice-cooker-like
mechanism should work. \\
\addlinespace[6pt]
\textbf{Experience} &
Intended use qualities, sensory outcomes, or affective qualities, such as
portability, ease of use, safety, or humour. &
Require evidence of the intended quality; a handle alone does not establish
portability. S4's canvas annotation names portability, while C5's speech
connects carrying and on-the-go use to design features. \\
\end{longtable}
\endgroup

The same submission can express form, function, and experience through
different or related evidence. On-canvas annotations count as canvas
evidence in both conditions. Functional detail or design rationale is not
automatically coded as Experience: the evidence must express an intended
use quality, sensory outcome, or affective quality. Participant-level
summaries count each participant once per category expressed across their
available submissions.

\clearpage
\subsection{Human Follow-up}
\label{sec:codebook-followup}

\textbf{Guiding question:} How does the participant respond to AI Interprets?
The unit is the participant's follow-up, read together with the preceding
AI interpretation. The framework draws on prior work on human--AI
co-creation~\cite{davis2026human}. Codes distinguish how the response acts on
the interpretation, rather than whether it contains a request for an image.
The paired examples are illustrated in Figure~\ref{fig:response_codes}.

\begingroup
\small
\setlength{\tabcolsep}{5pt}
\renewcommand{\arraystretch}{1.15}
\begin{longtable}{@{}p{.15\linewidth}p{.40\linewidth}p{.40\linewidth}@{}}
\caption{Human Follow-up codebook.}\label{tab:codebook-followup}\\
\toprule
Code & Definition and scope & Distinction and example \\
\midrule
\endfirsthead
\toprule Code & Definition and scope & Distinction and example \\
\midrule
\endhead
\bottomrule
\endfoot
\textbf{Execute} &
Accept the current interpretation and request its execution or
visualisation. &
Proceed without adding a design specification or restating a particular
design intention. Example: C1 asks the AI to make the visualisation. \\
\addlinespace[6pt]
\textbf{Reinforce} &
Confirm or emphasise an intention already present in the interpretation
without substantially changing its direction. &
Unlike Execute, the response explicitly reinforces design content.
Example: after the AI identifies a portable toaster, C5 requests an image
showing how the design would be portable. \\
\addlinespace[6pt]
\textbf{Elaborate} &
Add details, constraints, or refinements to the current design direction. &
Extend the interpretation without repairing its basic premise.
Example: after the AI recognises the toaster, S10 asks to add flower
designs. \\
\addlinespace[6pt]
\textbf{Reframe} &
Repair or redirect an interpretation that does not capture the intended
object, concept, or design premise. &
Distinguish repair of a misunderstanding from adding a new detail.
Example: after the AI identifies a DVD player, S10 clarifies that it is
a toaster whose bread exits sideways onto a plate. \\
\addlinespace[6pt]
\textbf{Ideate} &
Invite AI contributions to generate, compare, evaluate, critique, or
explore design possibilities. &
Open a design question rather than supply a specific refinement.
Example: C5 asks how to make unique modifications to the toaster. \\
\end{longtable}
\endgroup

A request to generate an image can accompany different responses: the
distinction depends on whether the participant simply proceeds, reinforces
an intention, adds detail, repairs the premise, or invites exploration.
Subsequent chatbot exchanges provide evidence for full-idea Creative Agency
coding rather than being treated as part of this immediate follow-up.

\clearpage
\subsection{Creative Agency}
\label{sec:codebook-agency}

\textbf{Guiding question:} How do human and AI contributions shape the full
idea? The unit is a participant-confirmed idea, including its canvas
submissions, AI interpretations, participant responses, and generated
outputs. One pattern is assigned to each idea. The framework draws on prior
work on human--AI collaboration~\cite{shi2026taxonomy,li2026ai}.

\begingroup
\small
\setlength{\tabcolsep}{5pt}
\renewcommand{\arraystretch}{1.15}
\begin{longtable}{@{}p{.15\linewidth}p{.40\linewidth}p{.40\linewidth}@{}}
\caption{Creative Agency codebook.}\label{tab:codebook-agency}\\
\toprule
Code & Definition and scope & Distinction and example \\
\midrule
\endfirsthead
\toprule Code & Definition and scope & Distinction and example \\
\midrule
\endhead
\bottomrule
\endfoot
\textbf{Human-only} &
The idea develops without incorporating AI contributions. The participant
forms and develops the design through their own contributions. &
AI output may be available without being incorporated. Unlike Human-led,
this pattern does not involve selective incorporation of AI input. \\
\addlinespace[6pt]
\textbf{Human-led} &
The participant primarily directs the idea while selectively incorporating
AI input to realise or refine the intended design. &
The participant retains the main design direction. S10 corrects the bus
interpretation and specifies toaster features before visualisation. \\
\addlinespace[6pt]
\textbf{Co-evolution} &
The idea develops reciprocally through human and AI contributions, with
participants and AI building on each other's contributions across exchanges. &
Generating an image alone does not establish reciprocal development.
C6 invites suggestions for integrating a freezer and toaster, then directs
their incorporation into a prototype. \\
\addlinespace[6pt]
\textbf{AI-led} &
The idea is primarily shaped by AI contributions, with the participant
mainly selecting or accepting the direction proposed by AI. &
AI contributions drive the design direction rather than primarily realising
a participant-specified direction or participating in reciprocal development.
Producing the final image alone does not establish this pattern. \\
\end{longtable}
\endgroup

The classification concerns development across the full idea sequence, not a
single follow-up or who produced the final image. Figure~\ref{fig:trajectory-example}
illustrates Human-led and Co-evolution examples. All four patterns were
retained in coding. Figure~\ref{fig:creative-agency} summarises participants
by their most frequent pattern, retaining ties; Human-only and AI-led were
not the most frequent pattern for any participant.


\end{document}